\documentclass[a4paper,fleqn,usenatbib,useAMS]{mnras}
\usepackage{graphicx}	
\usepackage{amsmath}	
\usepackage{amssymb}	
\usepackage{multicol}        
\usepackage{bm}		
\usepackage{pdflscape}	
\usepackage{array}
\usepackage{fancyvrb}
\usepackage{enumitem}

\usepackage[T1]{fontenc}
\usepackage{ae,aecompl}
\usepackage{txfonts}
\usepackage{threeparttable}
\usepackage{breakurl}
\usepackage{longtable}
\usepackage{booktabs}
\usepackage{xcolor}

\newcommand{\teff}{$T_{\rm{eff}}$}
\newcommand{\logg}{$\rm{\log g}$}
\newcommand{\feh}{[Fe/H]}

\title[Automated Stellar Spectral Classification]
{Application of Convolutional Neural Networks for Stellar Spectral Classification}
\author[Sharma et al.]{Kaushal Sharma$^{1}$ \thanks{E-mail:kaushals@iucaa.in}, Ajit Kembhavi$^{1}$, Aniruddha Kembhavi$^{2}$, T. Sivarani$^{3}$, \and
Sheelu Abraham$^{4}$, and Kaustubh Vaghmare$^{1}$\\
\vspace{1pt}\\
$^{1}$Inter University Centre for Astronomy and Astrophysics, Pune - 411007, India\\
$^{2}$Allen Institute for Artificial Intelligence, Seattle, USA\\
$^{3}$Indian Institute of Astrophysics, Bengaluru, India\\
$^{4}$Marthoma College, Chungathara - 679334, Nilambur, Kerala, India
}

\begin{document}

\date{Received on ; Accepted on }

\pagerange{\pageref{firstpage}--\pageref{lastpage}} \pubyear{2018}

\maketitle

\label{firstpage}

\begin{abstract}

Due to the ever-expanding volume of observed spectroscopic data from surveys such as SDSS and LAMOST, it has become important to apply artificial intelligence (AI) techniques for analysing stellar spectra to solve spectral classification and regression problems like the determination of stellar atmospheric parameters \teff, \logg{}, and \feh{}. We propose an automated approach for the classification of stellar spectra in the optical region using Convolutional Neural Networks. Traditional machine learning (ML) methods with ``shallow'' architecture (usually up to 2 hidden layers) have been trained for these purposes in the past. However, deep learning methods with a larger number of hidden layers allow the use of finer details in the spectrum which results in improved accuracy and better generalisation. Studying finer spectral signatures also enables us to determine accurate differential stellar parameters and find rare objects. We examine various machine and deep learning algorithms like Artificial Neural Networks (ANN), Random Forest (RF), and Convolutional Neural Network (CNN) to classify stellar spectra using the Jacoby Atlas, ELODIE and MILES spectral libraries as training samples. We test the performance of the trained networks on the Indo-U.S. Library of Coud{\'e} Feed Stellar Spectra (CFLIB). We show that using convolutional neural networks, we are able to lower the error up to 1.23 spectral sub-classes as compared to that of 2 sub-classes achieved in the past studies with ML approach. We further apply the trained model to classify stellar spectra retrieved from the SDSS database with SNR>20.

\end{abstract}


\section{Introduction}\label{sec:intro}

The Harvard scheme of stellar spectral classification divides stellar spectra into seven main classes following a temperature sequence: O, B, A, F, G, K and M, where O (\teff{}\,$>$\, 30,000\,K) represents the hottest and M represents the coolest (2200 - 3700 K) stars. Each main class label is followed by 10 subclasses ranging from 0 to 9 indicating decreasing temperature within that class. For example, a B0 type star would be hotter than a B9 type star but cooler than an O9 type star. An addition to the above classification scheme was proposed by Morgan and Keenan (the MK classification system) where each spectrum can also be classified into six primary luminosity classes: Ia, Ib, II, III, IV, and V. Luminosity class labels are appended to the main class and subclass (e.g. B9Ia). These labels are indicators of the stellar surface gravity, `Ia' denoting luminous giants (having least values of \logg{}) and `V' denoting dwarfs (having the maximum value of surface gravity) and hence add another physical property of the star to the classification system. A comprehensive description of the stellar spectral classes is provided in \citet{Graybook2009} and in a review article by \citet{Giridhar2010}.

MK Stellar Spectral Classification has conventionally been done by human experts through the visual inspection of each spectrum. With the arrival of modern computational capabilities and introduction of machine learning (ML) algorithms, the problem has been addressed in a more robust and less subjective manner where techniques such as $\chi^2$-minimisation, Artificial Neural Network (ANN), and Principal Component Analysis (PCA) have been applied \citep{Gulati1994,Bailer-Jones1998,Singh1998} to classify stellar spectra. Using ANN on the full spectrum in the optical region ($\sim$3600\,-\,7400 \AA), \citet{Gulati1994} achieved an average error of $\sim$2 spectral subclasses, whereas \citet{Singh1998} implemented ANN with PCA and obtained similar error of $\sim$2 subclasses but with reduced dimensionality. \citet{Bailer-Jones1998} demonstrated the application of PCA along with a `committee' of neural networks' to assign spectral and luminosity classes over the spectral types B2-M7 and luminosity classes III, IV, and V. They used stellar spectra from the Michigan Spectral Survey \citep{Houk1994} in the wavelength range 3800\,-\,5200 \AA{} and obtained an error of 1.1 subclasses in the regression mode. In the probabilistic mode of classification, they obtained an error of 2.09 subclasses. \citet{Manteiga2009} proposed a classifier, STARMIND, which tries to mimic the human reasoning for stellar classification by using a knowledge base comprising of spectral features computed from template spectra. The authors deployed an expert system (ES) which uses this knowledge base and applies fuzzy logic for classifying spectra with an overall accuracy of 82.7\% at the main-class level. \citet{Gray2014} developed a program named MKCLASS, which classifies stellar spectra (3800\,-\,5600 \AA) based on a comparison between the input spectrum and MK standard spectra using a human-like approach. The program is written in ``C'' language and not only assigns the standard MK classes but also looks for chemical peculiarities and other classes from modern classification system like white-dwarfs, Wolf-Rayet stars etc. In the MK classification system, they obtain an error of 0.6 subclass for the temperature sequence and 0.5 for luminosity class on spectra with S/N$\gtrapprox$100. 
Some studies \citep[e.g.][]{Navarro2012, Kesseli2017} used a set of spectral indices (line ratios, fluxes, etc.) rather than the full spectrum to classify stellar spectra with error of lower than two spectral subtypes. \citet{Kesseli2017} report an error of 1.5 subclasses using their classifier, PyHammer, on stellar spectra with R$\approx$2000 from Baryon Oscillation Spectroscopic Survey \citep[BOSS;][]{Dawson2013}. Outside the optical region, \citet{Vieira1995} and \citet{Weaver1997} have demonstrated the application of ANN for spectral classification in the ultraviolet and near-infrared (NIR) regions with accuracies of 1.1 and 0.5 subclasses respectively.

In this work, we apply shallow neural networks (NNs) as well as deep convolutional neural networks to study optical stellar spectra in detail and investigate whether using deeper networks of convolutional layers can significantly reduce the error and accuracy achieved in the stellar spectral classification. Deep learning frameworks \citep{Hinton2006,Bengio2009,Zeiler2013} have been used in the astronomical domain for various applications like galaxy morphology prediction \citep{Dieleman2015}, classification of variable stars based on their light curves \citep{Mahabal2017}, estimating atmospheric parameters using stellar spectra \citep{Fabbro2018}, detecting bar structures in galaxy images \citep{Abraham2018}, classifying galaxy morphologies at radio wavelengths \citep{Wu2019} etc. However, all these problems require a large sample for supervised training of the network. For the problem of stellar spectral classification, we do not have a dataset with more than 1500 stellar spectra with accurate classification made through independent studies. Empirical spectral stellar libraries (SSL) such as ELODIE \citep{Prugniel2001,Prugniel2007} CFLIB \citep{Valdes2004} and MILES \citep{miles2006} have been compiled by different groups to provide stellar spectra with wide coverage of the parameter space in the optical region of the electromagnetic spectrum. These libraries can be utilized as the training set for the classification using machine learning techniques. But the sample size remains a constraint for training a neural network with deeper layers. We address this problem with a class of machine-learning techniques called \textit{semi-supervised learning}, where pre-training of a network is performed on a large dataset which has the same or similar characteristics as the dataset of interest but with no identified spectral classes. Since the pre-training stage does not use classification labels for input features, the process is one of unsupervised learning. The pre-training phase adjusts the network weights close to their optimal value for a specific type of dataset. In the next step, a new model is created containing the first set of layers taken from the pre-trained model, concatenated with new layers which end up in a classification layer with the number of output nodes equal to the number of classes. This new model is fine-tuned using a labelled dataset, which can be smaller in size, for the supervised classification of the sample. For the unsupervised training of the network, we use a deep convolutional autoencoder with stellar spectra taken from Sloan Digital Sky Survey \citep[SDSS,][]{SDSSDR13} and use empirical SSL having labelled stellar spectra for the supervised training phase. For independent testing of the trained model, we classify stellar spectra from CFLIB which are not included in the training sample. The final model is applied to $\sim$ 48000 SDSS spectra, spectral classes are assigned and compared with the classification label provided in the SDSS database. We release the spectral and luminosity classifications assigned to the CFLIB and SDSS spectra as an online catalogue.

The paper is structured as follows. In Section~\ref{sec:data}, we provide an overview of spectral databases that have been used in this work. Different stellar spectral libraries (SSL) have different characteristics which require some pre-processing to use all spectra together as a single larger training set. We describe these steps in Section~\ref{sec:pre-processing}. In Section~\ref{sec:techniques}, we illustrate two different approaches for stellar classification, namely machine learning (ML) and deep learning (DL), define a set of metrics used for testing the performance of classification models and present the results of their application. We also apply these methods for the luminosity classification. We discuss some CFLIB sources which are misclassified by our classification model followed by a few diagnostics to establish the validity range and limitation of the current model in Section~\ref{sec:misclassification}. Following that, in Section~\ref{sec:sdss_application}, we present the application of the trained classification model on SDSS data and compare classes assigned by our model and the SDSS pipeline. Finally, we summarize this work and describe the future scope and prospects in Section~\ref{sec:results}.

\section{Data}\label{sec:data}

Developing any supervised classification model using machine/Deep learning techniques requires a standard dataset containing a large number of examples with known, reliable, and homogeneous classification which is used for the training and testing purposes. 
But due to unavailability of such an ideal sample of MK standards which is large enough for ML/DL applications, we depend upon merging spectra from different stellar spectral archives to prepare a larger training set.
Since our primary goal is to classify large spectral databases such as SDSS/LAMOST, we select stellar libraries which have the maximum overlap in the wavelength coverage with the target databases. With this primary constraint, we choose four spectral libraries, Jacoby-Hunter-Christian (JHC) Atlas \citep{Jacoby1984}, ELODIE \citep[V3.1]{Prugniel2007}, MILES \citep{miles2006}, and CFLIB \citep{Valdes2004} for training and testing of the ML/DL classifiers. We list the characteristic features of these databases in Table~\ref{tab:spectroscopic_databases}. 

\begin{figure*}
\centering
\includegraphics[scale=0.28]{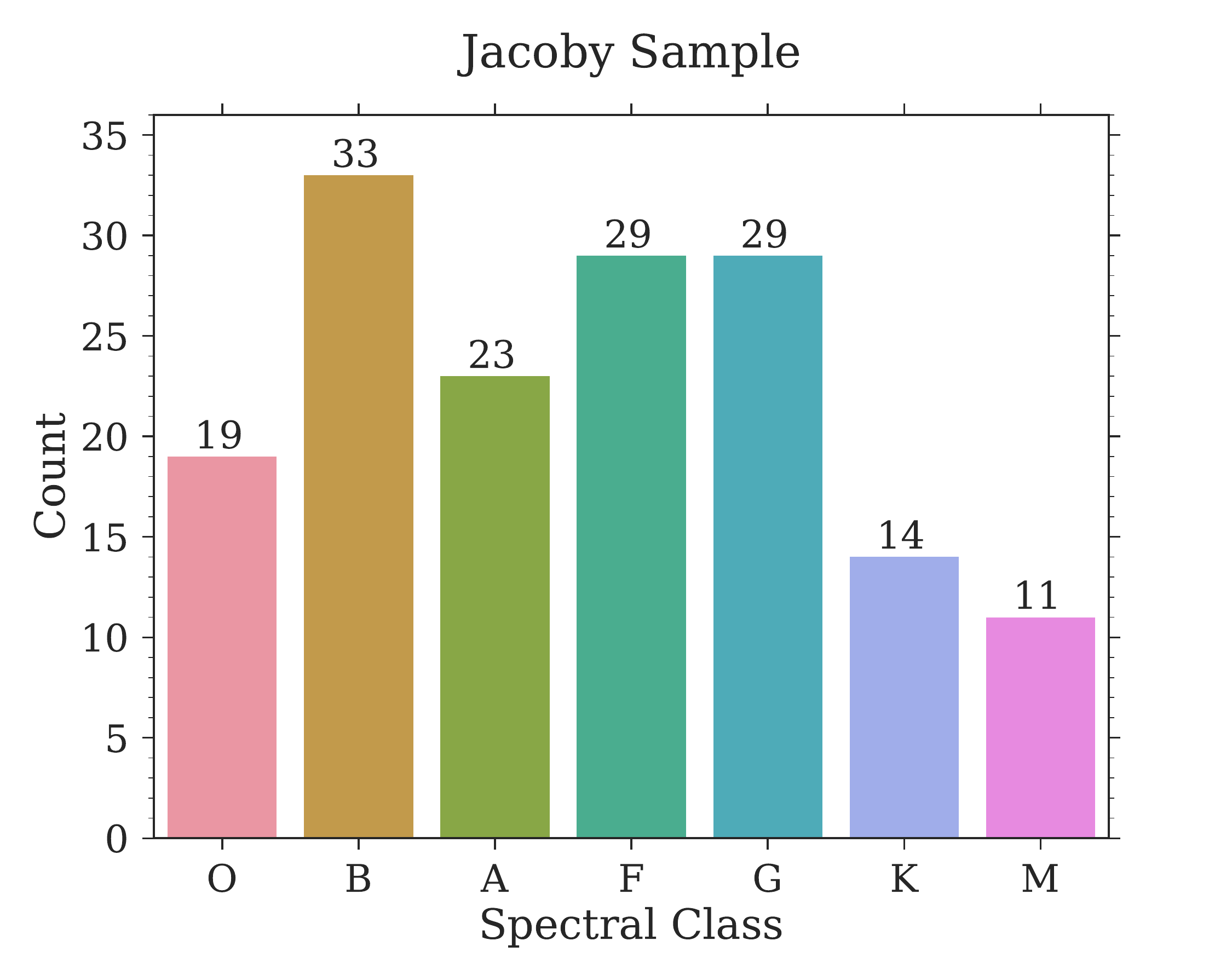}
\includegraphics[scale=0.28]{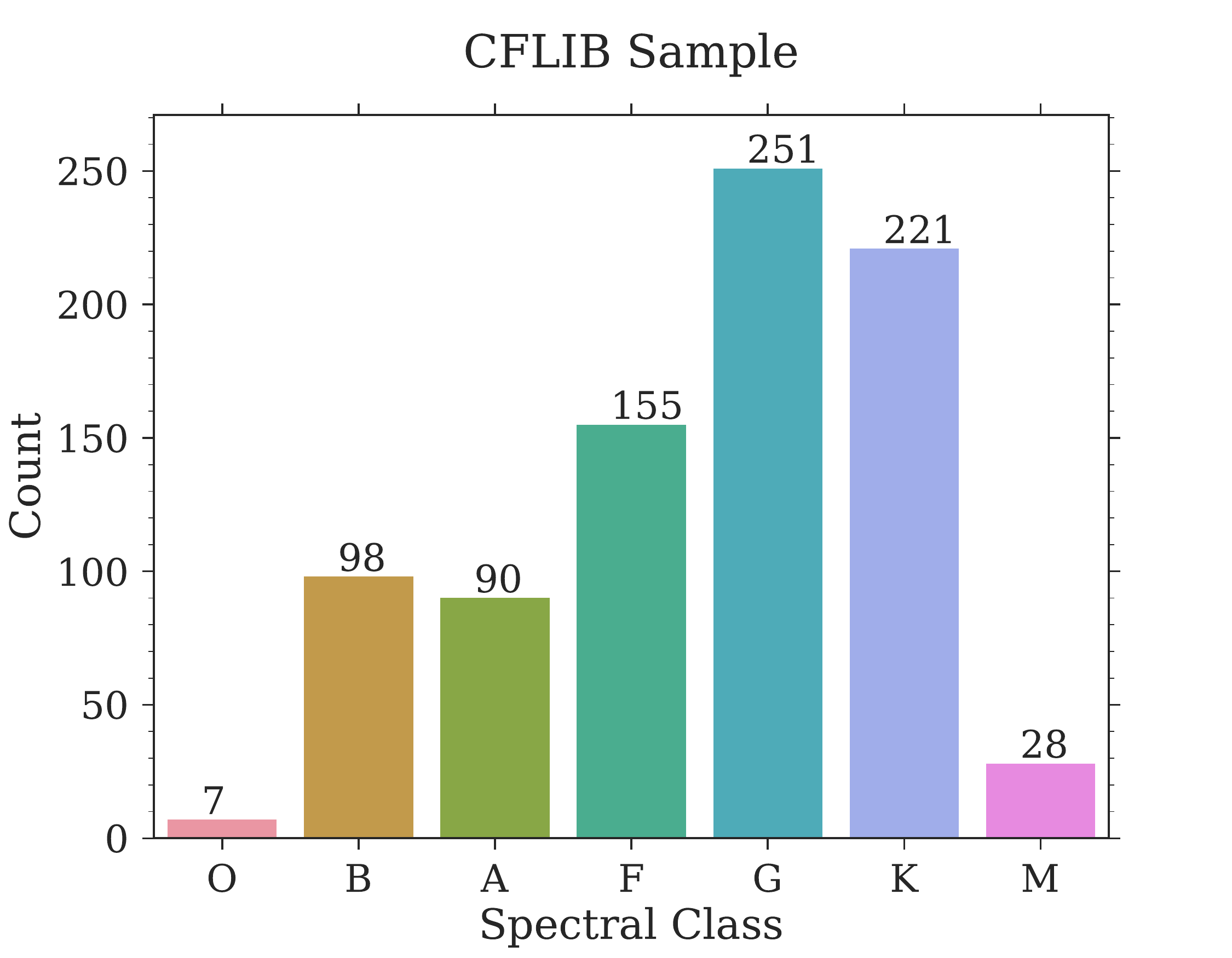}
\includegraphics[scale=0.28]{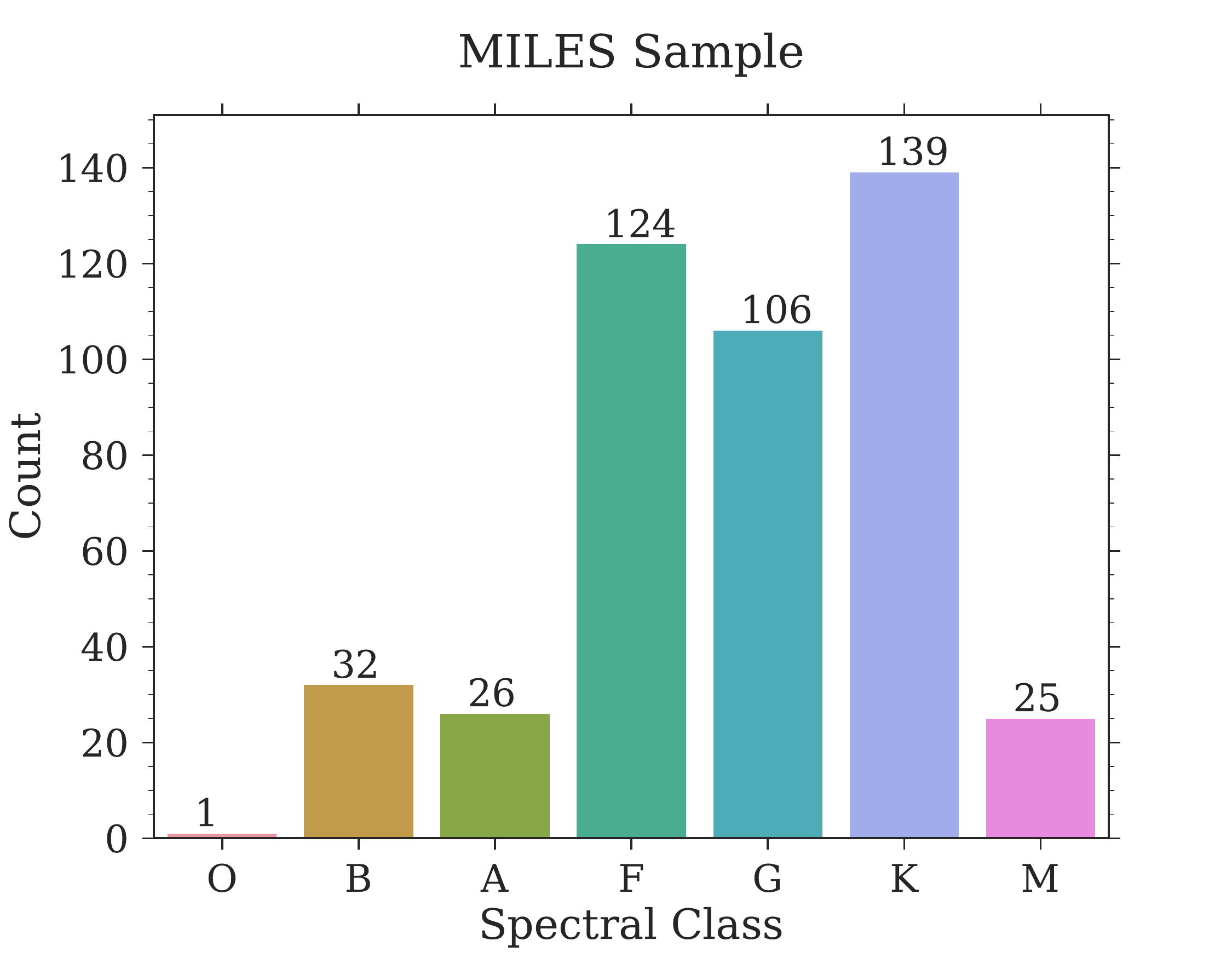}
\includegraphics[scale=0.28]{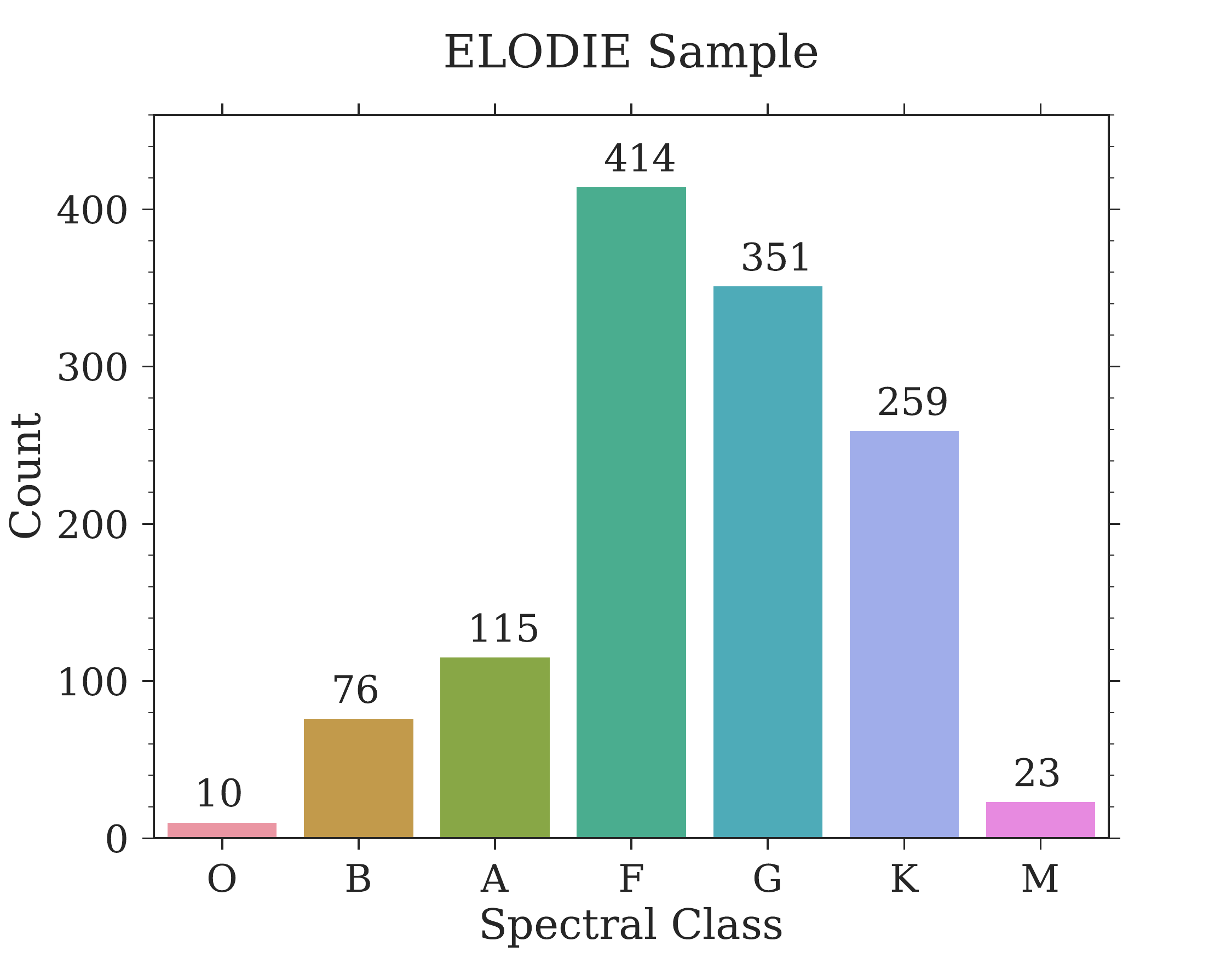}
\caption{Distribution over main spectral classes for the stars from various spectroscopic databases. The distribution is shown only for those stars which are selected for this study and not for all the spectra in these libraries.}
\label{fig:database_distribution}
\end{figure*}

For the spectral class for each of the sample spectra, we primarily rely on the classification provided in the source catalog but we also query the identifier in SIMBAD database and check the availability of the stellar classification from other studies to guard against any error in the adopted MK class. In most of the cases, the classes in the parent compilation are found to be consistent with the literature. For the small number of cases (less than 1\%) where we find a discrepancy, we visually inspect the spectrum in question and assign the class.
We present the distribution of spectra used in this study over the temperature classes in Fig.~\ref{fig:database_distribution}. Since we use CFLIB for an independent examination of the trained classifer, any common star between CFLIB and other libraries is excluded from the other libraries and is retained only in CFLIB.

We note that combining different libraries to prepare a larger training sample can potentially introduce inhomogeneities in the training set for two main reasons. First, different spectral databases have different instrumental characteristics, and second, the classification methods employed for classifying these databases are not the same and might suffer from systematic differences. To minimize the effects due to first source of inhomogeneity, we pass each spectrum through a pre-processing stage. For the latter one, we make an internal consistency test. These are discussed in more detail in Sec.~\ref{sec:pre-processing}.
We provide the relevant details about each spectral library used in this work below.

\begin{table*}
\caption{Characteristics of different spectral databases used in this study.}
\centering
\begin{tabular}{lccll}
\hline\hline
Database                & Selected Sample/ & $\lambda$ Coverage & FWHM Resolution (\AA{}) & Reference           \\
                        & Number of Stars  &      (\AA{})       & (R\,=\,$\lambda/\Delta\lambda$)  &            \\
\hline                                    
JHC Atlas               &     158/161      & 3510\,-\,7427      & 4.50 (R $\sim$ 1200 )  & \citet{Jacoby1984}   \\
ELODIE.3.1              &     1248/1959    & 3900\,-\,6800      & 0.57 (R $\sim$ 10000)  & \citet{Prugniel2007} \\
CFLIB                   &     850/1273     & 3460\,-\,9464      & 1.00 (R $\sim$ 5000)   & \citet{Valdes2004}   \\
MILES                   &     453/985      & 3536\,-\,7410      & 2.56 (R $\sim$ 2000)   & \citet{miles2006}    \\
\hline
\end{tabular}\label{tab:spectroscopic_databases}
\end{table*}

\subsection{Jacoby-Hunter-Christian Atlas}

The JHC Atlas\footnote{\url{ftp://ftp.stsci.edu/cdbs/grid/jacobi/}} consists of flux-calibrated spectra of 161 stars with spectral types ranging from O5-M7 and covering all luminosity classes (I, II, III, IV, and V), but dominated by type I, III and V. The spectral types in the library have been assigned through the visual inspection of each feature \citep{Jaschek1978}. We select 158 stars having solar composition. Of the remaining three stars, two are metal poor (HD 94028, SAO 102986 with atlas number 153 and 154) and one (SAO 81292 atlas number 155) is a double or multiple star system with emission lines in the spectrum. The main class distribution for the library is shown in the upper left panel of Fig.~\ref{fig:database_distribution}. Luminosity classes from I-V contain 45, 8, 44, 4, and 57 spectra respectively. The library spans the wavelength range of 3510\,-7427 \AA{} at a resolution of 4.5 \AA.

\subsection{Indo-U.S. Library of Coud{\'e} Feed Stellar Spectra (CFLIB)}

For CFLIB, sometimes also referred to as Indo-US library, spectra of 1273 stars were obtained at resolution of $\sim$1 \AA{} FWHM over the region $\lambda\lambda$ 3460\,-\,9464 \AA{}. The stars were selected to provide a broad coverage of the parameter (\teff, \logg{} and \feh{}) space. To create the library, observations were taken over a period of eight years in five different grism settings to obtain the desired wavelength coverage. This resulted in 6917 individual spectra for 1273 stars. In some observing runs, the unavailability of exact same grating settings caused gaps in the wavelength coverage. For our analysis, we chose only those stars for which the full wavelength coverage is available or the missing region is small (< 50 \AA) and/or lies beyond the optical region. This gives us a sample of 850 stars. There are two types of spectral classification provided in \citet{Valdes2004}:

\begin{enumerate}[leftmargin=0.6\parindent]
\item spectral type from the literature, and
\item spectral type from \citet{Pickles1998} empirical library used for defining the continuum shape.
\end{enumerate}

We noticed that the CFLIB spectral types from the literature are not complete and sometimes luminosity classes are not provided. Therefore, we adopt the latter for preparing our database. To check the consistency of the two classification schemes, we compared the two series of spectral types (wherever spectral type from the literature is available) and find that the two agree within 0.5 subclasses. The main-class distribution for the library is presented in the right panel of the upper row in Fig.~\ref{fig:database_distribution}. Luminosity classes I-V comprise of 46, 16, 349, 80 and, 359 spectra respectively.

\subsection{MILES Spectral Library}

The Medium-resolution Isaac Newton Telescope Library of Empirical Spectra \citep[MILES,][]{miles2006} is a compilation of 985 flux-calibrated stellar spectra at FWHM resolution of 2.56 \AA{} \citep[R $\sim$ 2000,][]{Prugniel2011} over the region of $\lambda\lambda$ 3536\,-\,7410 \AA{} and normalized at 5550 \AA. For compiling the spectral classes, we searched the SIMBAD database and found that the spectral and luminosity classes are available for 453 MILES stars, which we adopt in our compilation. The remaining 532 stars are not included in our sample. Class-wise distribution for selected MILES stars is shown in the lower left panel of Fig.~\ref{fig:database_distribution}. The number of stars belonging to five luminosity classes are 1, 5, 164, 60, and 223 respectively.

\subsection{ELODIE 3.1 Spectral Library}

The ELODIE library\footnote{\url{http://perso.astrophy.u-bordeaux.fr/CSoubiran/elodie_library.html}} (named after the ELODIE echelle spectrograph attached to 1.93m telescope at Observatoire de Haute-Provence) is another high-resolution spectroscopic database which provides the stellar spectra at two resolving powers: R $\sim$ 42000 where the flux has been normalized to the continuum, and R $\sim$ 10000 by Gaussian broadening where the flux has been calibrated in physical units using the photometry from the Tycho catalogue. The first version of the library was presented in \citet{Prugniel2001} which contained 908 spectra of 709 stars with the wavelength coverage of 4100\,-\,6800 \AA. We use the latest release of the library \citep[V3.1]{Prugniel2007} containing 1962 spectra of 1388 stars with improved data reduction/flux calibration and extended wavelength coverage towards bluer region starting from 3900 \AA{} to 6800 \AA.

In the ELODIE library, individual spectra are available in FITS format with the header containing the relevant information such as atmospheric parameters and assigned spectral class, with the latter being adopted from the Hipparcos INCA database which is a compilation of astrophysical measurements.
Like the other sources of spectral types, here too in some cases the luminosity class is missing. We cross-identify such cases with SIMBAD and adopt the luminosity classes from there. This results in 1248 unique spectra of 886 individual stars whose class-wise distribution is shown in lower right panel of Fig.~\ref{fig:database_distribution}. The five luminosity classes (I-V) contain 22, 11, 253, 125, and 837 spectra respectively.

\section{Pre-processing and Internal consistency}\label{sec:pre-processing}

Any ML/DL algorithm requires a large amount of data for training a model and performance of the trained model greatly depends upon the quality of the training sample. A small training sample (of the order of a few hundred) can result in an over-fitted model. In case of NNs, a smaller training set constrains the depth of the network in terms of the number of hidden layers and is not suitable for the deeper architectures. Therefore, to have a large training set which has significant representation of each class in the training sample, we combine spectra from the JHC Atlas, MILES and ELODIE library and use the combined sample as the training set. This sample has 1859 spectra over different spectral types. 

As mentioned in Sec.~\ref{sec:data} and evident from Table~\ref{tab:spectroscopic_databases}, these databases have different characteristics (FWHM resolution, flux scales, wavelength coverage etc.), and we need to homogenize the data before using it as a training sample. For that, we adopt a 3-step pre-processing procedure:

\begin{enumerate}[leftmargin=0.6\parindent]
\item We degrade all the spectra to the same resolution (poorest among all the chosen databases) by convolving with a Gaussian kernel of suitable width,
\item normalize flux values in all spectra to unity at 5550 \AA, and
\item use a cubic spline function to interpolate available flux values to a common grid of wavelength points.
\end{enumerate}

We processed 158, 1248, 850 and 453 spectra respectively from JHC Atlas, ELODIE, CFLIB and MILES libraries, described in Sec.~\ref{sec:data} through these steps and brought all spectra to resolution of 4.5 \AA{} by convolving with a Gaussian whose FWHM equals $\sqrt{(4.5)^2-R_{\textrm{original}}^2}$. Since the common wavelength coverage for these databases is 3900\,-\,6800 \AA, we interpolate the spectra in this range with a step size of 1 \AA. We keep the pre-processed CFLIB sample for testing the trained network. The above steps then result in a training matrix (from JHC Atlas, ELODIE, and MILES) of size $1859\times2900$ and test matrix (from CFLIB) of size $850\times2900$, with  each spectrum labelled with a spectral class and a luminosity class. We also check the signal-to-noise ratio (SNR) for the spectra included in our study and find that nearly 70\% of spectra in the training set and 65\% spectra in the test set have the SNR in the range 15-100. About 20\% spectra in both the sets have SNR in the range 100-200.

After the pre-processing, all spectra from a given subclass should ideally look similar, but there is bound to be some dispersion which we have to keep in mind while preparing the training sample. The origin of this dispersion could be attributed to remnant instrumental effects in the spectra after pre-processing stage or to different classification methodologies adopted for different libraries. We assess this effect by estimating the internal dispersion for each spectral subclass. For this, we first compute the mean spectrum denoted by $\boldsymbol{\mu}$: $ \{ \mu_1, \mu_2, ...., \mu_j,...., \mu_m \}$, for every subclass in our dataset where $m$ represents the number of wavelength points. For $n$ spectra in a subclass, the flux values at $j^{th}$ wavelength point of the mean spectrum is given by:
\begin{eqnarray}\label{eq:mean_flux}
\mu_j & = & \frac{1}{n}\sum_{i=1}^{n}F_{ij},~~~~~1\le\,j\,\le2900,
\end{eqnarray}
where $F_{ij}$ is the flux value for the $i^{th}$ spectrum at the $j^{th}$ point. After computing the mean spectrum, we determine the dispersion of the $i^{th}$ spectrum, $\sigma_i$, with respect to the mean spectrum using
\begin{eqnarray}\label{eq:indv_int_dispersion}
\sigma_i & = & \sqrt{\frac{1}{m} \sum_{j=1}^{m} (F_{ij} - \mu_j)^2}.
\end{eqnarray}
Finally, the internal dispersion $\sigma$ for a subclass is the mean of the dispersion values for the subclass:
\begin{eqnarray}\label{eq:int_dispersion}
\sigma & = & \frac{1}{n} \sum_{i=1}^{n} \sigma_i.
\end{eqnarray}

\begin{figure}
\centering
\includegraphics[scale=0.28]{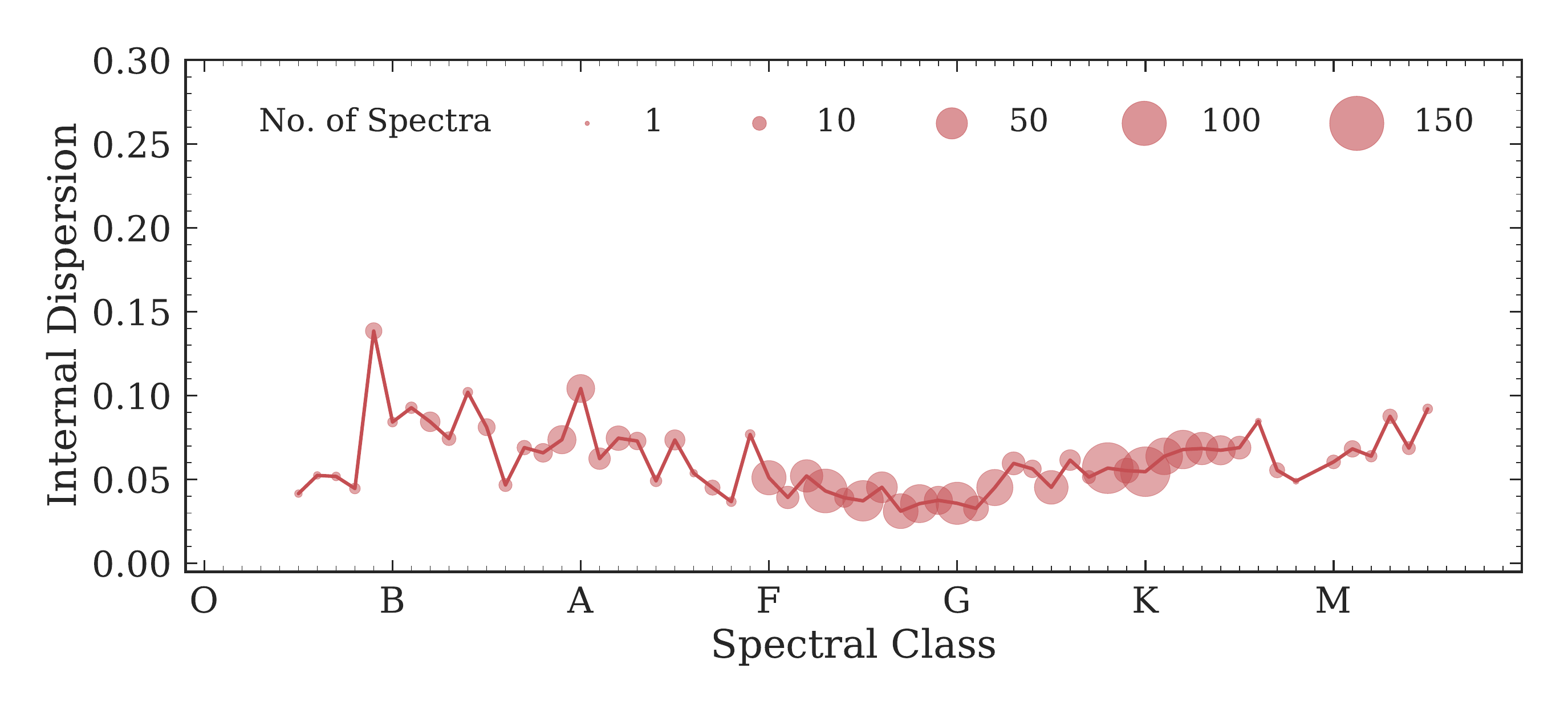}
\caption{Internal dispersion in the training set for various spectral subclasses. Those subclasses which are not present in the training set (e.g. O4, K9) are not shown in the figure. Size of the circle denotes the population of a subclass as indicated in the legend.}
\label{fig:int_dispersion}
\end{figure}

For a subclass having a single spectrum, the internal dispersion is considered to be zero. Fig.~\ref{fig:int_dispersion} shows the internal dispersion for various subclasses present in the training set. The average value of internal dispersion for the training set comes out to be 5 percent in flux units after removing 3$\sigma$ outliers. We feel this is reasonable for the spectra to serve as a good training set. It is worth noting here that ML/DL algorithms with properly chosen cost functions are robust to a small fraction of examples being supplied with wrong labels as the weights of the network are adjusted by the features present in the majority examples and are not significantly influenced by small contamination in the class labels. We verify this aspect of our model by supplying erroneous classes for a limited number of spectra and finding that the performance metrics do not show significant decline for the test set. We discuss this in greater detail in Sec.~\ref{sec:sensitivity}.

\section{Classification Methodology}\label{sec:techniques}

As mentioned in Sec.~\ref{sec:intro}, there have been various attempts in the past to apply machine learning methods to the problem of stellar spectral classification. \citet{Gulati1994} uses ANN with two hidden layers, each layer having 64 neurons and obtained an error of two-subclasses in their classification scheme. \citet{Bailer-Jones1998} use a committee of neural networks and obtain 2.09 subclass error in the classification mode. \citet{Navarro2012} set-up a two-stage classification system which is trained on spectral indices and outputs the spectral and luminosity class. Three networks in the first stage contain two hidden layers with 50 neurons in each layer. Similarly, the second stage networks also have two hidden layers with 30-50 neurons in these layers. In all these works, a training set of rather small size was used. With the availability of a bigger training set now, we attempt the problem of spectral classification using traditional ML algorithms as well as deep-learning architectures like Convolutional Neural Network (CNN) with autoencoder architecture.

For implementation of machine learning techniques, we use various classification and regression tools available in the \texttt{scikit-learn}\footnote{\url{http://scikit-learn.org/stable/index.html}} library \citep{scikit-learn}. For the deep-learning models, we use Keras API with TensorFlow \citep{tensorflow2015} as the backend. The following subsections describes the details of the various networks and their performance.

\subsection{Machine Learning: Methods and Results}\label{sec:ml_methods}

For any supervised ML algorithm, a class label (for a classification task, where the output is a categorical variable) or a number (for a regression task, where the output is a quantitative variable of continuous nature) is assigned to each input feature vector (pre-processed spectrum in our case). This labelled dataset is used for training the ML model. Based on the nature of the output variable, the problem of spectral classification can be addressed as follows:

\begin{itemize}
\item [-] as a traditional multi-label, multi-class classification problem where each spectrum is assigned two labels, one corresponding to the spectral class out of 70 classes and another out of five luminosity classes according to MK system. Though there are six main luminosity classes identified in the MK system, as mentioned in Sec.~\ref{sec:intro}, due to a lack of enough training spectra in classes Ia and Ib, we combine the two into a single class I.
\item [-] as a regression problem where each spectrum is assigned a number based on its spectral and luminosity class and the network is trained to predict these numbers which are decoded back to their original class. This reduces the dimensionality of the problem from two (spectral and luminosity class for each input feature) to one. 
\end{itemize}

For classification, we start with a basic neural network architecture in Keras which contains one input layer, two hidden layers and an output layer with the number of nodes equal to the number of spectral classes i.e. 70.
At this stage, the classification is limited to the 70 spectral subclasses and luminosity classifcation is performed later. First the spectral classes are integer-encoded with e.g. O0 being 0, F5 being 35, and M9 being 69. Some classification frameworks like \textit{Keras} require the labels to be binarized and have the same dimension as the number of classes. In such a binarized representation, class O0 will be represented as `1' followed by 69 zeros [1,0,0,0,0,0,....]. Similarly classes O1 and M5 would be represented as [0,1,0,0,0,....] and [....0,0,0,1,0,0,0,0] respectively. This step results in a matrix of $n\times70$ dimension, where n is the number of spectra. This process is called one-hot encoding, which we perform for integer-encoding the classes. To determine an optimized set of most significant training parameters, namely batch size, number of neurons in each layer, best performing activation function, optimizer and initializer, we perform a hyper-parameter optimization over the following grid of parameters:

\begin{Verbatim}[fontsize=\small]
 - batch_size = [60, 80, 100, 120]
 - neurons1 = [32, 64, 128]
 - neurons2 = [32, 64, 128]
 - optimizer = [`RMSprop', `Adagrad', `Adadelta', `Adam',
                `Adamax', `Nadam']
 - activation = [`softmax', `softplus', `softsign',
                 `relu', `tanh', `sigmoid', 
                 `hard_sigmoid', `linear']
 - initializer = [`uniform', `lecun_uniform', `normal',
                `glorot_normal', `glorot_uniform',
                `he_normal', `he_uniform']
\end{Verbatim}
The above grid results in $4\times3\times3\times6\times8\times7\,=\,12096$ combinations. Running the algorithm over all possible settings is computationally expensive.
Therefore, rather than sweeping through the whole hyperparameter space, we choose a random combination and fit the model. The architecture is `cross-validated' (as discussed below) five times and the accuracy (the fraction of total spectra assigned correctly to their original class) is evaluated and saved. The model is then dropped for further training on the next random combination of the parameters. This process is repeated 100 times, each time with a random combination of hyperparameters and each combination is validated 5 times. A total of 500 fits are performed which take $\sim$ 8.5 hours on a GPU node equipped with a single Tesla P100-PCIE-16GB GPU card and 3584 cores. Finally, the training parameters corresponding to the best accuracy are returned and are further used in the final model.

\begin{figure}
\centering
\includegraphics[scale=0.55]{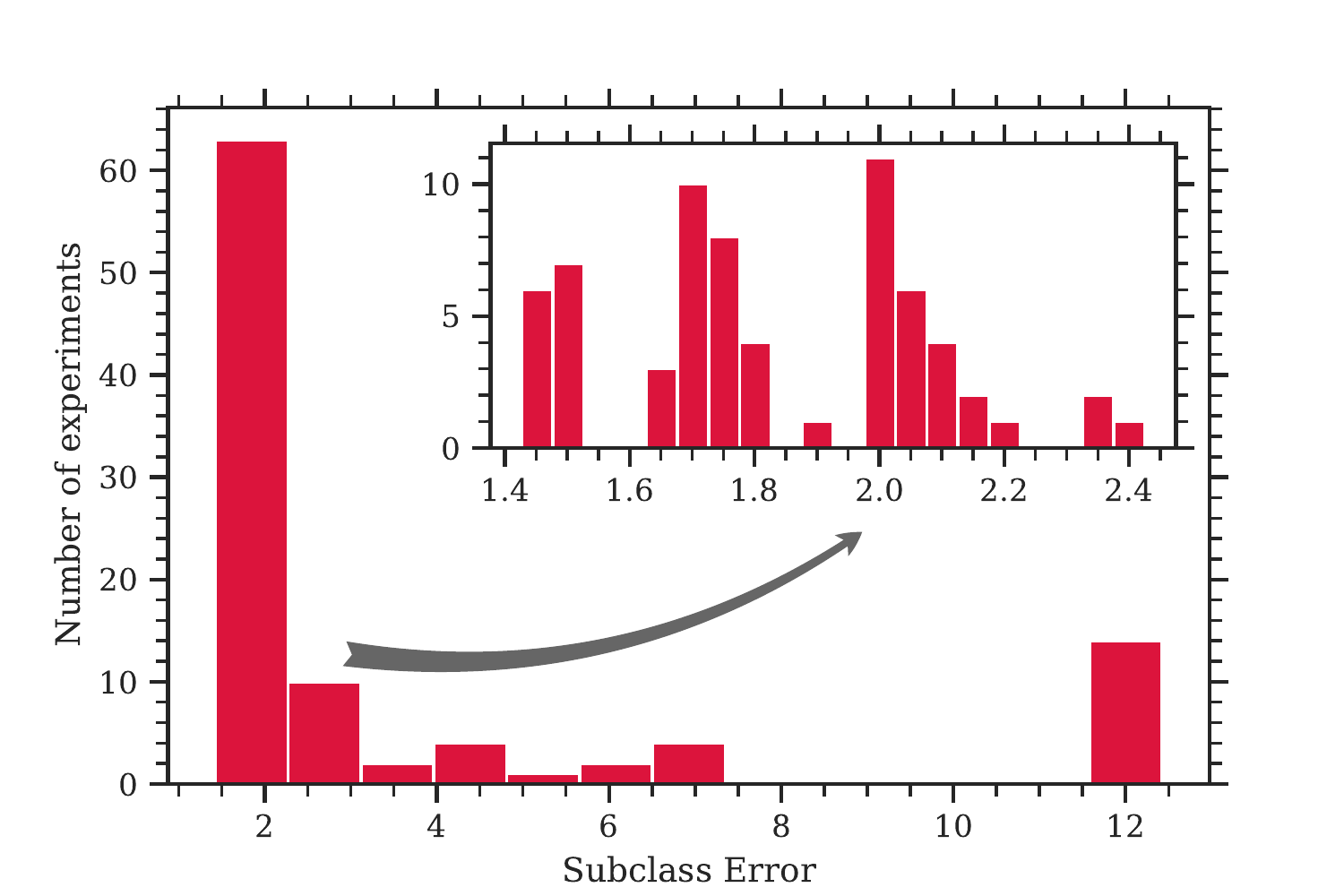}
\caption{Distribution of subclass error as scoring parameter for the random search for hyperparameter optimization. The inset plot shows the distribution in the range of 1.4\,-\,2.5 subclass error.}
\label{fig:seta_randsearch_stability}
\end{figure}

The best random search model returns subclass error ($\sigma$, introduced later in this section) of 1.43 subclasses for batch size and epochs equal to 120 and 1000 respectively, and `ReLU' \citep[Rectified Linear Units;][]{relu2010} activation with `adam' optimizer \citep{adam} and `uniform' initializer. The best grid model uses 32 and 64 neurons in the first and second hidden layers respectively. To test the sensitivity of the model on the hyperparameter selection, we plot the distribution of subclass errors as the scoring parameter for these 100 combinations. A uniform distribution will mean that our model is sensitive to the chosen hyperparameters, whereas a distribution tightly clustered around lower error values would indicate the robustness of the model to the hyperparameter settings. This distribution is presented in Fig.~\ref{fig:seta_randsearch_stability} and shows that most combinations result in lower subclass error, clustered in the range of 1.4-2.2 subclasses. Only a few combinations result in higher subclass error. This analysis also shows that we are not overfitting the model through hyperparameter selection.

By experimenting with a larger number of hidden layers and a larger number of neurons per layer, we find that the accuracy improves by less than 1\%. Moreover, increasing the number of neurons in hidden layers comes at the cost of a large number of trainable parameters which makes the learning model more susceptible to over fitting. Therefore, we limit ourselves to a model with two hidden layers with the first having 32 neurons and the second 64 neurons.

\begin{figure*}
\centering
\includegraphics[scale=0.55]{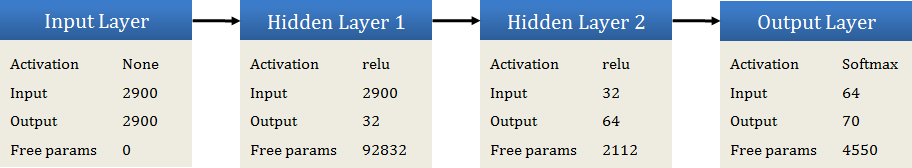}
\caption{Schematic diagram showing the architecture of the neural network classifier. Number of neurons and trainable parameters in each layer are indicated as `Output' and `Free params' respectively. Each layer uses `uniform' initializer to initialize the network weights.}
\label{fig:seta_nn_classifier}
\end{figure*}

Fot training the architecture, we use 85\% of randomly selected sources for training in each epoch and the remaining 15\% for validation. 
We then pass the spectra in batches of 120 through the network to minimize the loss function with `adam' optimizer. We choose categorical cross-entropy \citep{Goodfellow2016} loss-function, $H(p,q)$, which is minimized during the model training process to obtain the best model. It is a measure of the mismatch between the true probability distribution $p(x)$ and predicted probability distribution $q(x)$ and is given by:
\begin{eqnarray}\label{eq:loss_function}
H(p,q) = -\sum_x p(x)\log(q(x)).
\end{eqnarray}

\begin{figure*}
\centering
\includegraphics[scale=0.39]{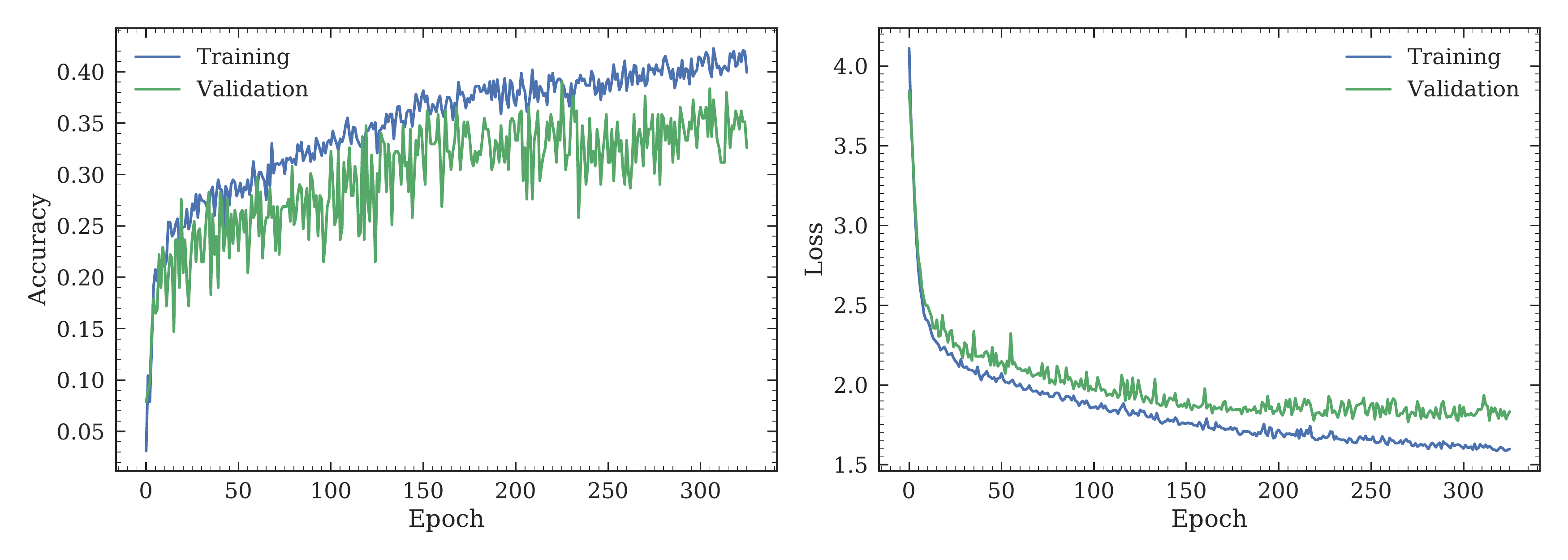}
\caption{Learning curve for the ANN model for training and validation samples as a function of number of epochs of learning. Model accuracy (sub-class) and loss function (categorical cross-entropy) tracked over different epochs are shown in the left and right panels respectively.}
\label{fig:seta_training_keras}
\end{figure*}

To evaluate the performance, we check the accuracy after each epoch. We use an upper limit of 1000 epochs for training the network, but to avoid over-fitting, we stop early if no significant improvement in the validation accuracy is achieved for 50 consecutive epochs. With this condition for early stopping, our model stops training after 325 epochs. Fig.~\ref{fig:seta_training_keras} shows the performance evolution of the model over the epochs of training.

We observe that conventional accuracy (the fraction of total spectra assigned correctly to their original class) is not the best measure to check the performance of a model in our case as the classification labels follow a particular sequence, while the conventional accuracy metric is completely opaque to whether the misclassification is `close' to its original class or far away. Therefore we use another metric which is the mean $\mu$ and standard deviation $\sigma$ of the difference between integer-encoded predicted and expected classes. $\mu$ and $\sigma$ denote the bias and the error respectively at the subclass level. We also measure the correspondence between the expected and predicted class by doing a linear regression analysis between the expected and predicted spectral codes as 
\begin{equation}\label{eq:linear_regression}
\qquad\qquad\qquad\qquad\boldsymbol{y_{\textrm{pred}}}=m\times\boldsymbol{y_{\textrm{expd}}}\,+\,c,
\end{equation}
where $\boldsymbol{y_{\textrm{pred}}}$, $\boldsymbol{y_{\textrm{expd}}}$ are the predicted and expected spectral codes respectively. $m$ and $c$ denote the \textit{slope} and \textit{intercept} of the linear fit. Another parameter that we use to evaluate the \textit{goodness} of prediction is R$^2$-Score \citep{Steel1960,Glantz1990,Draper1998}, also referred to as coefficient of determination, which indicates the proportion of variance in the predicted labels governed by the expected labels. 
R$^2$-Score can take the values in the range 0-1, and is defined as: 
\begin{eqnarray}\label{eq:r2_score}
R^2 = 1 - \frac{SS_{\textrm{res}}}{SS_{\textrm{tot}}},
\end{eqnarray}
where the sum of square of residuals $SS_{\textrm{res}}$ and the total sum of squares $SS_{\textrm{tot}}$ are defined as:
\begin{eqnarray}\nonumber
SS_{\textrm{res}}=\sum _{i}(y_{i}-f_{i})^{2}=\sum _{i}e_{i}^{2};\\\nonumber
SS_{\textrm{tot}}=\sum _{i}(y_{i}-{\bar {y}})^{2}.
\end{eqnarray}
Here $y_i$ and $f_i$ represent the expected and predicted labels and $\bar{y}$ denotes the mean of the expected labels. For a perfect classification model, the values of slope, intercept and R$^2$-Score should be equal to 1, 0, and 1 respectively with zero bias ($\mu$) and error ($\sigma$).

In the above training process, we randomly divide the training and validation set and shuffle the two sets after every epoch. But there is still a chance that the network is not trained to learn all the features in the dataset as some input vectors might never enter the training set and always end up in the validation set. In such a scenario, the trained model performs well for the training set but does not give the desired results on the unseen data. Such a model is considered to be unstable and might not capture all the relevant features from the training set. Therefore, we perform 10-fold cross-validation where the whole sample is shuffled randomly and divided into 10 subsets out of which one unique subset is held out for validation and the remaining subsets form the training sample. The model that we want to check for stability is then fitted on the training sample and accuracy is evaluated for the validation subset. This process is repeated 10 times by holding a different subset for validation each time. Finally, we compute the average of accuracy from each run which comes out to be $40.0\pm3.1$\%. This value is similar to the accuracy of the final model (maximum accuracy for training set in Fig.~\ref{fig:seta_training_keras}) with very small variation. This test confirms the stability of the neural net architecture for the training set. The subclass accuracy of 40\% translates to an error of $\sim$1.4 subclasses. We use the subclass accuracy just as an indicator of the model accuracy during the training stage; it is not used to evaluate the final performance of the ML/DL models developed in this work. 

\begin{figure}
\centering
\includegraphics[scale=0.39]{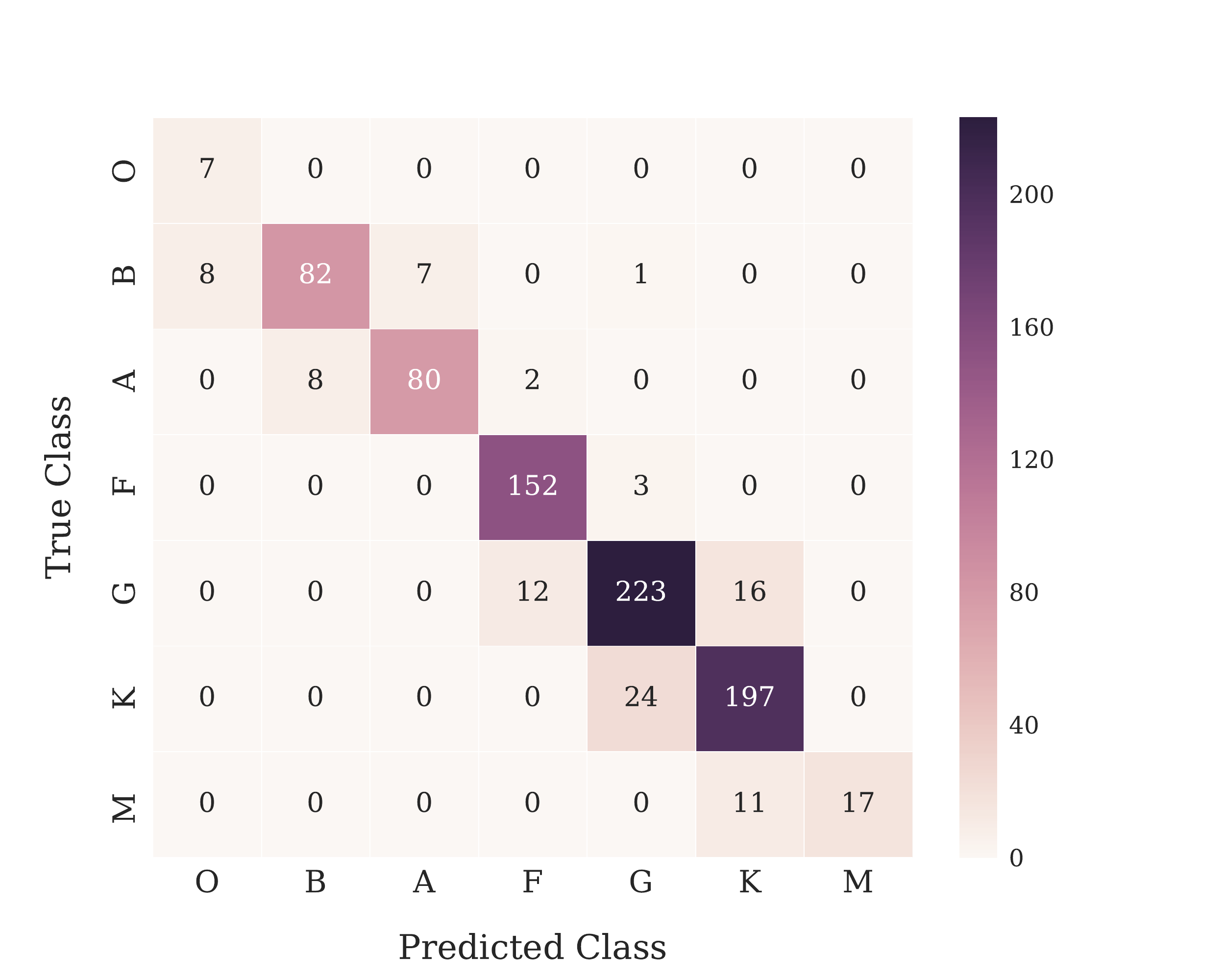}
\caption{Confusion matrix for the main spectral classes using Keras ANN model computed for the CFLIB library.}
\label{fig:seta_cm_keras}
\end{figure}

After training the classification model and cross-validating its stability, we test it on 850 spectra ($850\times2900$ matrix) from CFLIB and compare the predicted classes with the expected classification compiled from the literature. To examine the performance of the classification model we first check the confusion matrix (Fig.~\ref{fig:seta_cm_keras}), for the seven main spectral classes, which shows the distribution of sample spectra over the expected and predicted classes. 
In Fig.~\ref{fig:seta_cm_keras}, most of the spectra lie along the diagonal, which is an indication of a good classification model. Non-zero off-diagonal numbers indicate departure of the classification model from the ideal model. For example, 24 K type spectra are labelled as G type and 16 G type spectra are classified as K type by the classification model, so these classes are the most confusing for the model. But we realize that we should consider not the numbers but the fraction of spectra going off-diagonal as that truly characterizes the performance of the classification model. This is shown in Table~\ref{tab:seta_cm_keras} where we tabulate the precision, recall and F1-score (harmonic average of the precision and recall) for each of the main classes. 

Precision for a given class is the fraction of all spectra placed in it which have been correctly classified. Recall is the fraction of spectra of a given class correctly identified by the model. These two terms and the F1-score are defined as:

\begin{align}\label{eq:precision}
\qquad\qquad \textrm{Precision} & = \frac{\textrm{Correctly classified spectra in a class}}{\textrm{Total classified spectra in that class}},\\\label{eq:recall}
\qquad\qquad \textrm{Recall} & = \frac{\textrm{Correctly classified spectra in a class}}{\textrm{Total spectra in that class}},\\\label{eq:f1}
\qquad\qquad F1 & = 2\,\frac{\textrm{Precsion}\times \textrm{Recall}}{\textrm{Precsion} + \textrm{Recall}}.
\end{align}
For example, a total of 166 stars have been labelled as F type, while only 152 of these are truly F type. Thus the precision of 152/166\,=\,0.92. All seven O type stars in the test sample have been labelled correctly as O type, so the recall is 1.0 (7/7).

\begin{table}
\caption{Classification report for the main classes using Keras ANN.}
\centering
\begin{tabular}{lcccc}
\hline\hline
Class & Precision & Recall  & F1-score & Population \\\hline
     O & 0.47      & 1.00    & 0.64     & 7         \\
     B & 0.91      & 0.84    & 0.87     & 98        \\
     A & 0.92      & 0.89    & 0.90     & 90        \\
     F & 0.92      & 0.98    & 0.95     & 155       \\
     G & 0.89      & 0.89    & 0.89     & 251       \\
     K & 0.88      & 0.89    & 0.89     & 221       \\
     M & 1.00      & 0.61    & 0.76     & 28        \\
Total  & 0.90      & 0.89    & 0.89     & 850       \\\hline
\hline
\end{tabular}\label{tab:seta_cm_keras}
\end{table}

\begin{figure}
\centering
\includegraphics[scale=0.33]{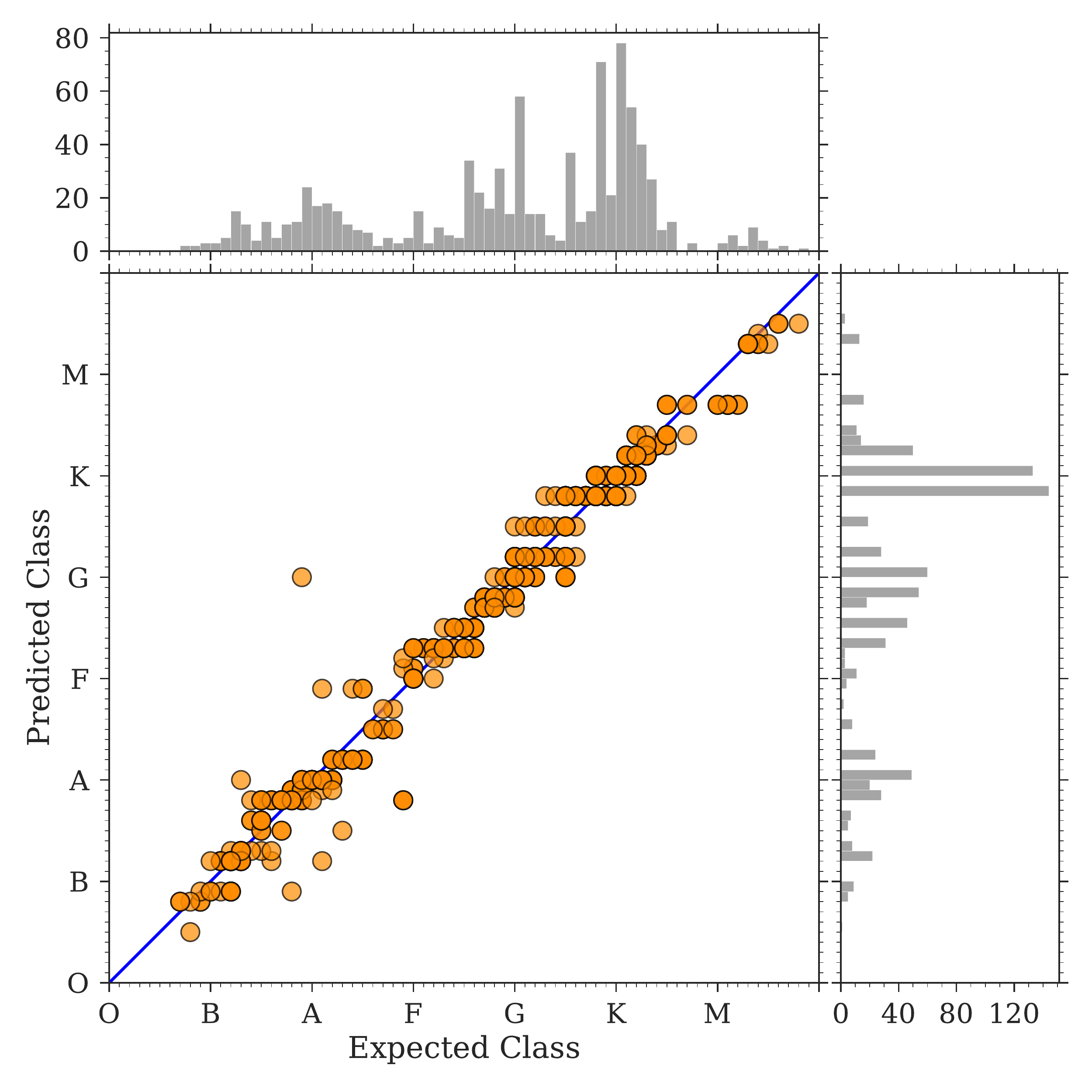}
\caption{Comparison between the predicted and expected classes using Keras ANN model. The histograms next to the top and right axes show the distribution of stars over the expected and predicted classes respectively.}
\label{fig:seta_keras_regression}
\end{figure}

For further assessment of the model, we compute the R$^2$ score (Eq.~\ref{eq:r2_score}) between the predicted and expected spectral codes which returns a value of 0.98. We also perform linear regression between the expected and predicted classes, to obtain $m$\,=\,0.99 and $c$\,=\,$0.67$. Fig.~\ref{fig:seta_keras_regression} highlights this comparison between the predicted and expected classes. 

To get a quantitative measure of the accuracy at the subclass level in the classification model, we 
calculate the mean $\mu$ and standard deviation $\sigma$ of the difference ($\boldsymbol{y_{\textrm{pred}}}$\,$-$\,$\boldsymbol{y_{\textrm{expd}}}$). We find $\mu\,=\,-0.27$, $\sigma\,=\,1.35$, which shows that there is, in effect, no systematic deviation in the predicted classes as compared to the expected classes and the classification is accurate up to 1.35 spectral subclass. 

To check other classification algorithms, we repeat the same steps on the training set using a multi-layer perceptron (MLP) classifier implemented in Python library \texttt{scikit-learn}. The number of hidden layers and neurons in each layer is kept the same (32, 64) and hyper-parameter optimization is performed over a grid of randomly selected activation function (logistic, relu, Tanh), solver (lbfgs, sgd, adam), $L2$ regularization parameter ($10^{-1}-10^{-7}$), batch size (60, 80, 100, 120), and learning rate (``constant'', ``invscaling'', ``adaptive''). The best accuracy is obtained for [activation = ``logistic'', solver =  ``lbfgs'', alpha = $10^{-1}$, batch size = 80, learning rate = ``adaptive'']. L-BFGS \citep[Limited-memory Broyden-Fletcher-Goldfarb-Shanno;][]{Nocedal1980, lbfgs2007} is a quasi-Newtonian method to update the weights after each training epoch. After scanning the parameter space grid for the combination resulting in the best accuracy, the dataset is divided into training and validation sets and the hyperparameter-optimized model is fitted to the training set. We obtained an accuracy of 41.2\% for the validation set. The cross-validation for this model returns an accuracy of 42.1\%$\pm$4.1\%. The trained model is tested on 850 spectra from CFLIB and the performance is evaluated on the basis of the same parameters as adopted for the Keras ANN classifier. Values of main-class accuracy, slope and intercept of the fitted linear regression model, $R^2$-score along with bias and error in the classes predicted are reported in Table~\ref{tab:classification_results}.

Another algorithm that we apply is the random-forest \citep[RF;][]{Breiman2001} classifier from scikit-learn. We perform a grid search over the RF algorithm hyperparameters and find the accuracy optimized hyperparameters: [n\_estimators = 600, min\_samples\_split = 2, min\_samples\_leaf = 2, max\_features = `sqrt', max\_depth = 110, bootstrap = False]. n\_estimators is the number of decision trees used in the Random Forest, min\_samples\_split is the minimum number of samples required to carry out a split for a given internal node, min\_samples\_leaf is the minimum number of samples allowed to be present in a leaf node, and max\_depth is the maximum depth to which a single tree is allowed to grow (also known as \textit{pruning}). max\_features is set to `sqrt' which means that any given tree uses a subset consisting of the square root of the total number of features of the training set. We use Gini impurity \citep{Gini1921,Breiman1984} as a split criterion for the tree branches and obtain a cross-validation score of 45.62\%$\pm$4.06\%. Prediction on the CFLIB sample using the RF classifier gives an accuracy of 40.8\%. Other performance metrics are provided in Table~\ref{tab:classification_results}.

\begin{table*}
\caption{Performance of different models on the CFLIB sample. Bias and error are computed by calculating the sigma-clipped statistics for (predicted class - expected class).}
\centering
\begin{tabular}{lccccrl}
\hline\hline
     Model          & Main-class Accuracy  &  Slope ($m$)  & Intercept ($c$)  & $R^2$-Score & Bias & Error \\
\hline
\multicolumn{7}{c}{Classification}\\\hline
  Keras ANN            &               89\%         & 0.99  & 0.67  &     0.98    &$-0.27$&    1.35   \\
  MLP Classifier       &               85\%         & 0.96  & 1.48  &     0.98    & 0.10  &    1.45   \\
  Random Forest        &               82\%         & 0.97  & 1.20  &     0.97    & 0.32  &    1.35   \\
  CNN with Autoencoder &               89\%         & 1.00  & 0.38  &     0.98    &$-0.14$&    1.23   \\\hline
\multicolumn{7}{c}{Regression}\\\hline
  MLP Regressor        &               85\%         & 0.99  & 0.35  &     0.98    &$-0.05$&    0.97    \\
  Random Forest        &               84\%         & 0.99  & 0.40  &     0.98    &$-0.07$&    1.16   \\
\hline
\end{tabular}\label{tab:classification_results}
\end{table*}

Out of the three classification algorithms we have trained, the neural network architecture using Keras performs best on test spectra from CFLIB.

In addition to conventionally assigning the 70 classes, we also try a regression approach which is motivated by the idea that the discrete spectral classes follow a continuous temperature sequence which decreases from O to M spectral type and also from 0 to 9 subclasses within each spectral class. Therefore, the spectral classes can be converted into a continuous variable according to the spectral encoding scheme described in \citet{Gulati1994} (refer to their Eq. 5). This scheme avoids the problem of a large number of classes. In this scheme, the classification problem is converted into a regression problem where each class is assigned a spectral code as follows:
\begin{equation}\label{eq:spectral_encoding}
\textrm{Spectral code} = 1000.0\times A1 + 100.0 \times A2 + (1.5 + 2\times A3),
\end{equation}
where $A1$ is the main spectral type of the star (O to M coded as 1 to 7), $A2$ is the subclass, ranging from 0 to 9, and $A3$ is the luminosity class (with the classes I-V coded as 0 to 4). In this scheme, a B9V type star will be labelled as 2909.5 and a K2III star as 6205.5. Each class here is an encoded number which changes the nature of the problem from classification to regression.

For obtaining a regression model which maps the flux values in the spectrum to their corresponding spectral code, we use MLP and RF regressor models. In order to obtain the best combination of model parameters, we perform a grid search in the hyper parameter space as earlier and adopt the ones which give the best prediction score. For assessing the prediction in this case, we use the R$^2$-Score (Eq.~\ref{eq:r2_score}).

For MLP regressor, we get the best R$^2$-Score\,=\,0.96 for a network with the configuration [2900:64:64:1], where there are 2900 input nodes, two hidden layers with 64 neurons in each layer and one output node for the spectral code. 
But using a configuration [2900:32:64:1] which has fewer neurons in the hidden layers, we get a similar R$^2$-Score of 0.95. 
Since a neural network with a larger number of neurons in the hidden layers has a larger number of trainable parameters, such configurations are susceptible to over-fitting. So, we finally choose the latter configuration and train the network. Each neuron in this network uses the ReLU activation function and for updating the model weights after each epoch, we use quasi-Newton procedure L-BFGS. We divide the whole training set into 85\% and 15\% ratios for training and validation respectively and train the model for 1000 epochs with batch size of 200. To avoid over-fitting of the model, we set an early stopping criterion on the R$^2$-Score of validation sample. If the validation score does not improve for 10 consecutive epochs, the training stops. The trained model returns a R$^2$-Score of 0.98 for the training sample.

For testing the trained network, we apply it to the 850 spectra from CFLIB and predict the spectral codes. The comparison between the predicted and expected spectral codes returns a R$^2$-Score of 0.98. For checking the classification accuracy over the main classes, we decode the spectral codes back to their original spectral and luminosity classes and find an accuracy of 85\% for the main classes. The values of $\mu$ and $\sigma$ for $\boldsymbol{y_{\textrm{pred}}}$\,$-$\,$\boldsymbol{y_{\textrm{expd}}}$ turns out to be $-5$ and 97 in code units respectively, indicating that the classification model has 1$\sigma$ error of 0.97 subclasses with no significant systematics in the prediction. 

According to the encoding scheme given in Eq~\ref{eq:spectral_encoding}, the luminosity classes are denoted by the last three characters in the coded number including the decimal point (e.g. `1.5' in `6201.5' represents the luminosity class I). For assigning the luminosity classes, we round off the predicted codes to one decimal place and extract the last three digits which are matched to their closest number out of [1.5, 3.5, 5.5, 7.5, 9.5] representing luminosity classes from I to V respectively. We compare the predicted luminosity classes with the original classes and get an accuracy of 36\%. The reason for poor accuracy for the luminosity classes is attributed to the lower weight given to the luminosity class as compared to spectral class in the spectral encoding scheme of Eq.~\ref{eq:spectral_encoding}. Therefore, the regression model does not work well for predicting spectral and luminosity classes together.

We also repeat the same regression exercise using another algorithm called Random Forest regressor with the default model parameters and get an accuracy of 84\% for the main class prediction on CFLIB. We get a 1$\sigma$ dispersion of 1.16 subclasses with $R^2$-score and slope of 0.98 and 0.99 respectively for the predicted vs true classification. For the luminosity classification, this model gives an accuracy of only $\sim$37\%. 

We find that for stellar spectral classification, neither regression model works for the luminosity classes and therefore we continue with the classification approach using Deep-learning models. The classification approach offers one more advantage, \textit{viz.} identifying stellar spectra of composite nature \citep{Bailer-Jones1998,Gray2014} as it operates in the probabilistic mode.

\subsection{Deep Learning: Method and Results}\label{sec:deep_net}

One major drawback with ML algorithms like ANN and RF is that these methods consider each input feature individually and do not take into account the sequence which these features follow. Such an approach is most appropriate when a system is described by a few properties and the output class does not depend upon the order in which these properties are fed to the ML algorithms. However, in our case, the input features are flux values which follow a sequence that characterizes the specific MK class. To address this issue, we use deep convolutional neural networks (DCNN), a class of deep-learning methods, which are suitable for tackling sequential data like a spectrum, time-series, audio pattern etc. 

CNN algorithms were originally designed to deal with image (2D/3D data) classification and recognition tasks where each neuron in the network learns a pattern in the input image using convolutional filters of varying sizes. Each neuron in a CNN connects only to a small portion of the input image as opposed to fully connected layers in traditional ANN, a concept that is referred to as \textit{local connectivity}. This helps in reducing the number of trainable parameters and in turn makes the computation more effective.

Since in our case the input features are 1-dimensional sequences, we use 1-D CNN architectures. For the application of any deep-learning algorithm, the major challenge is obtaining a large dataset with reliable labels for supervised training of the network. \citet{Fabbro2018} use a sample of 224,000 synthetic spectra for training a convolutional neural network \textit{StarNet}, which is used for determining the atmospheric parameters of stars using their spectra. In another application of deep neural nets, \citet{Pearson2018} generate a sample of $\sim$311,000 light curves showing transiting and non-transiting behaviour. The network is trained on this sample to learn how to detect Earth-like planets from time-series photometry using the transit method. In such cases, the sample size is more than adequate for deep learning methods. The labelled sample in our case is relatively small, which leads us to consider a semi-supervised deep learning algorithm.

\begin{figure}
\centering
\includegraphics[scale=0.15]{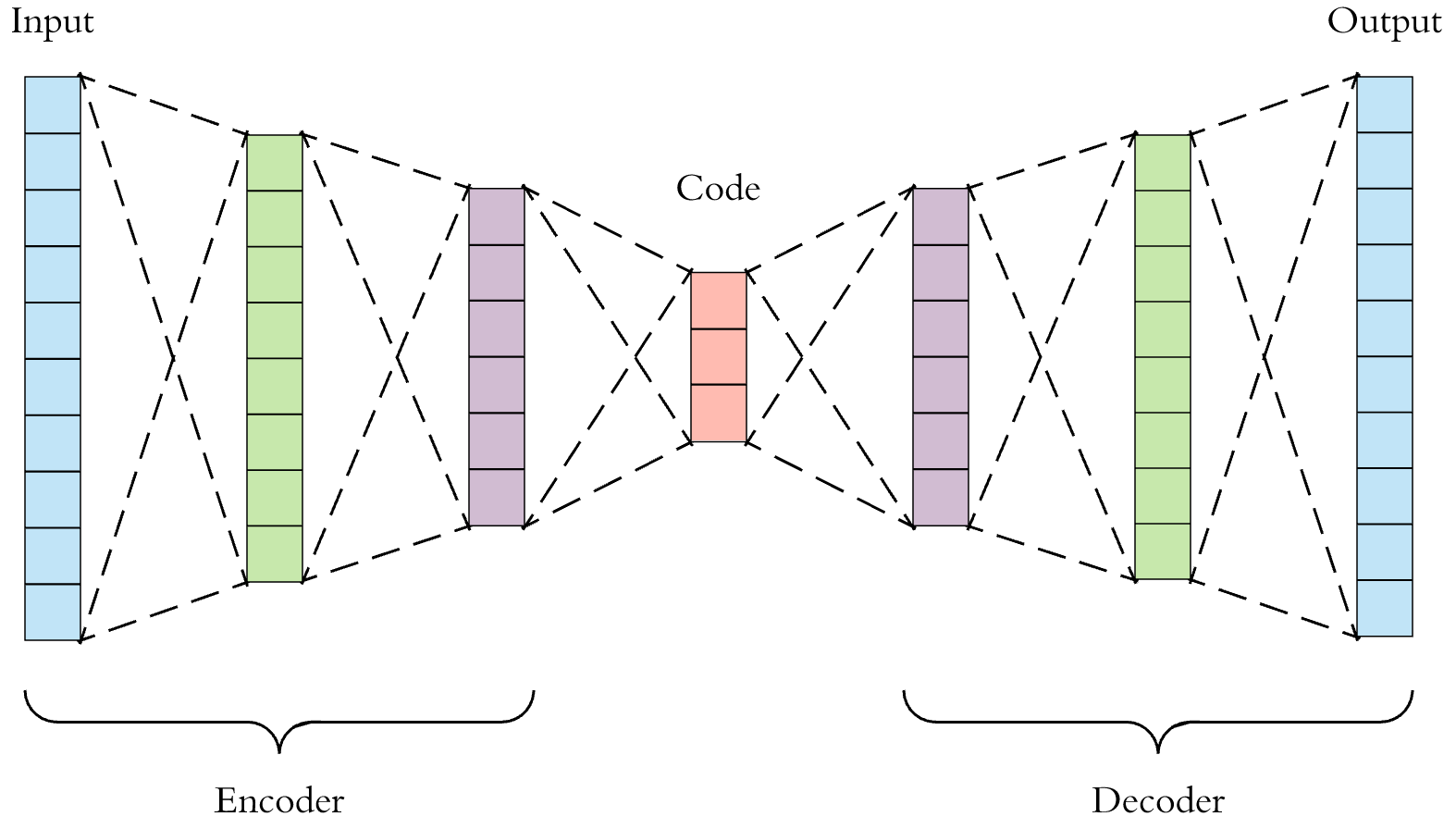}
\caption{Schematic diagram for the encoding and decoding layers of an auto-encoder architecture. ``Code'' represents the input data in a reduced-dimensional space. Image Credit: \url{https://towardsdatascience.com/applied-deep-learning-part-3-autoencoders-1c083af4d798}.}
\label{fig:ae_schematic}
\end{figure}

For semi-supervised learning on stellar spectra, we start with an autoencoder, which is an example of an unsupervised learning technique. The basic principle behind autoencoders is ``back-propagation using input features as output'' where the input features are mapped to themselves and hence do not require labels corresponding to input attributes. The training is performed to let the network learn the reconstruction of input features. The mapping between input and output nodes takes place through two sets of layers where one set, called the encoding layers, converts the input features into a compressed representation and another set, the decoding layers, maps the compressed form back to the input features. A schematic diagram for autoencoders is shown in Fig~\ref{fig:ae_schematic}. Autoencoders were introduced by \citet{Rumelhart1986} where the authors demonstrated that error back-propagation leads to a solution in almost every case, provided the problem is solvable. Later, the auto-encoder framework \citep{Hinton1994} was developed as denoising autoencoders \citep{Vincent2008} and dimensionality reduction tools, which may be considered to be a non-linear counterpart of PCA \citep{Hinton2006}. 

\begin{figure*}
\centering
\includegraphics[scale=0.25]{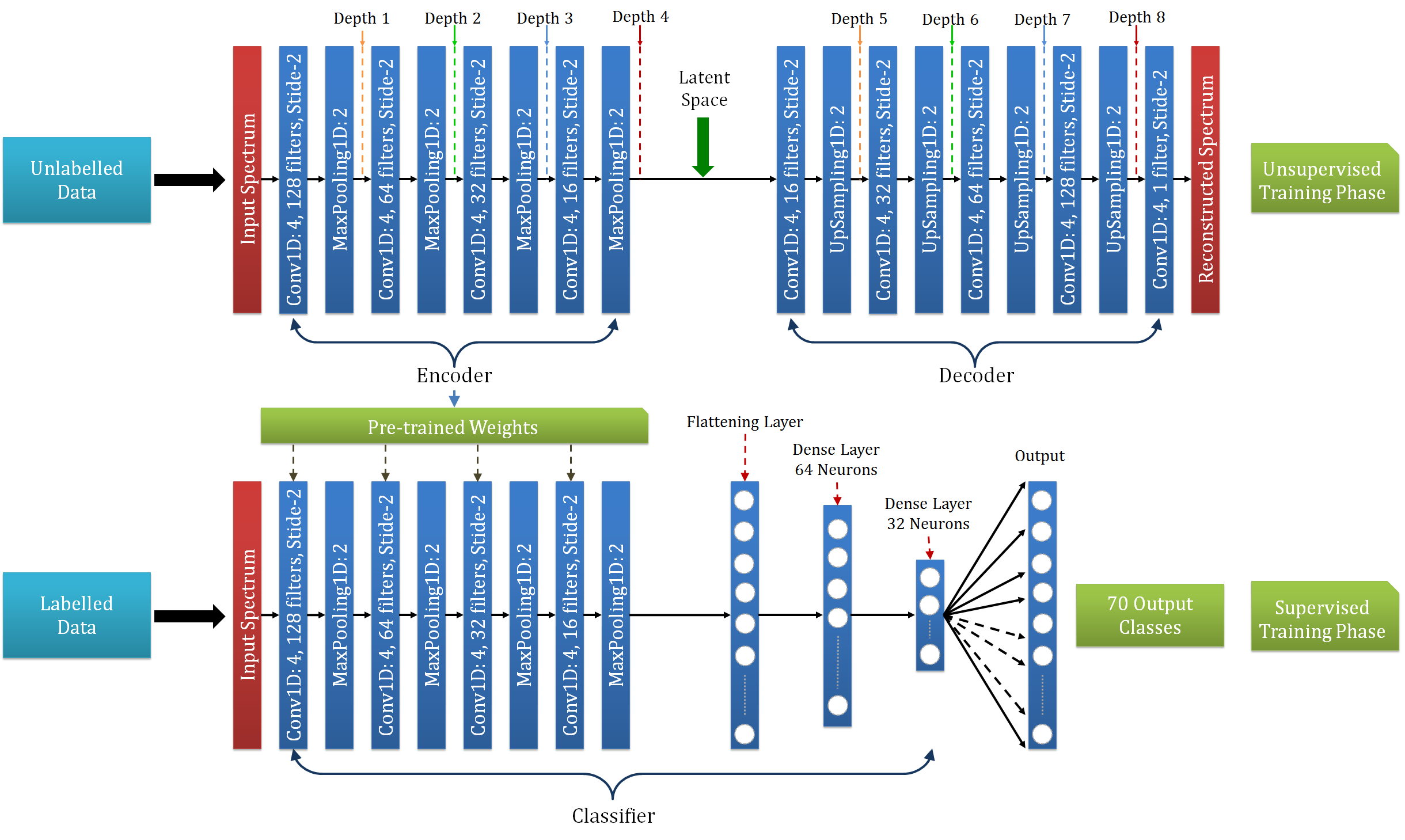}
\caption{Architecture of the semi-supervised 1-D CNN classification model using autoencoder. Encoder and decoder blocks together constitute the autoencoder part, shown in the upper half of the diagram. Training of the autoencoder part requires unlabelled data without known classes and therefore this phase is referred to as unsupervised training. This is the pre-training phase where the weights of the encoding and decoding layers are adjusted. After that, in the fine-tuning phase, the encoding layers with pre-trained weights are appended with flattening layers. Encoding and flattening layers together form the classification model which is supplied with the labelled data and outputs the predicted classes. This phase of the model training is referred to as supervised training because this requires data with known classes.}
\label{fig:ae_architecture}
\end{figure*}

To convert an autoencoder architecture into a semi-supervised classifier, we first train the autoencoder network on $\sim$60000 stellar spectra taken from the SDSS. After training, we create a separate model where (1) the decoding layers are omitted and (2) the tuned encoding layers are appended with an extra flattening layer and a couple of dense layers with random initial weights. The new model is trained on the labelled training set. With this supervised training, the encoding layers are fine-tuned and the weights are re-adjusted to classify the stellar spectra. The full architecture of this semi-supervised approach is presented in Fig~\ref{fig:ae_architecture} where the upper half shows the unsupervised training using the autoencoder and lower half shows the supervised fine-tuning part.

For the unsupervised training of the autoencoder model, we query the SDSS DR13 database via SQL using CasJobs\footnote{\url{http://skyserver.sdss.org/CasJobs/}} to retrieve a list of stellar spectra with a signal-to-noise ratio (SNR) greater than 20 with all warning flags set to 0. This query results in 61,627 spectra. Each spectrum is trimmed between 3900\,-\,6800 $\AA$ and resampled with 1 $\AA$ step. This provides 2900 wavelength points and corresponding flux values for each spectrum as the sample. The complete sample is divided into training, validation, and testing set with 50000, 5000 and 6624 spectra respectively. Out of the 2900 flux values, only the first 2896 (3900\,-\,6796 $\AA$) are supplied to the network to make it consistent with the final encoded layer. Each layer in the autoencoder model uses ReLU activation.

\begin{figure*}
\centering
\includegraphics[scale=0.4]{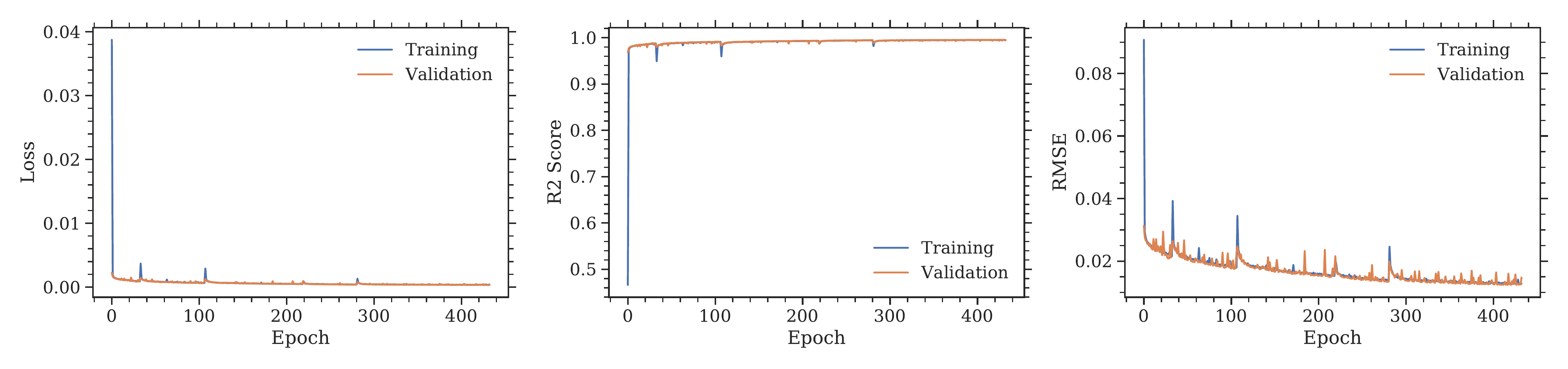}
\caption{Loss function, R$^2$-score and root mean squared error as a function of number of epochs of learning using 1-D CNN model.}
\label{fig:autoenc_training}
\end{figure*}

For training the model, we use a batch size of 256 with maximum epochs of 1000. The Adam optimizer is used for optimizing the model learning, and the mean squared logarithmic error between the input and the predicted output is used as the loss function. At every epoch, we compute the root mean square error and R$^2$-score to evaluate the performance of the model. The learning curve is shown in Fig.~\ref{fig:autoenc_training}.

\begin{figure}
\centering
\includegraphics[scale=0.32]{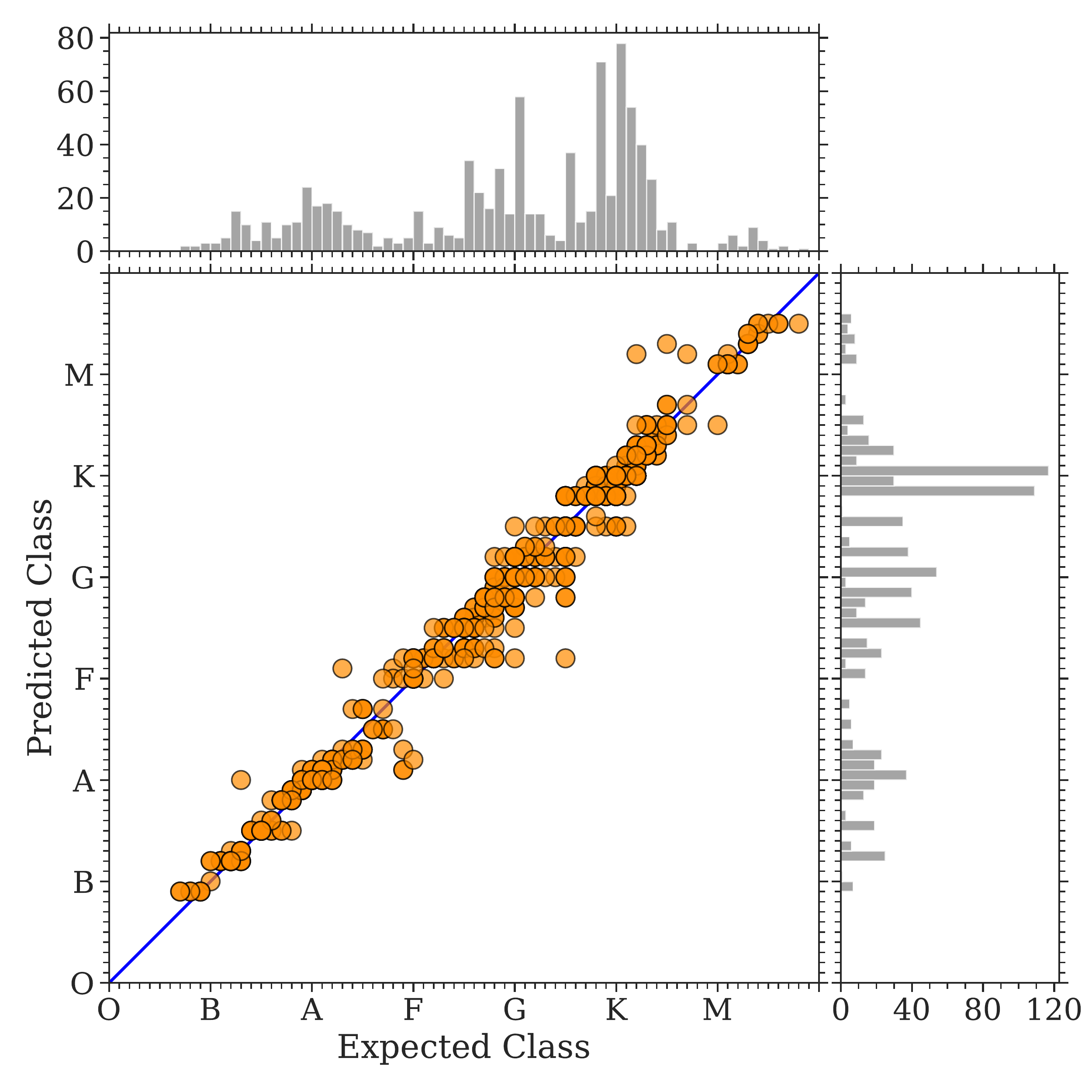}
\caption{Predicted classes as compared to expected classes using 1-D CNN classification model with autoencoder. The histograms next to the top and right axes show the distribution of stars over the expected and predicted classes respectively.}
\label{fig:ae_seta_classification}
\end{figure}

\begin{figure}
\centering
\includegraphics[scale=0.39]{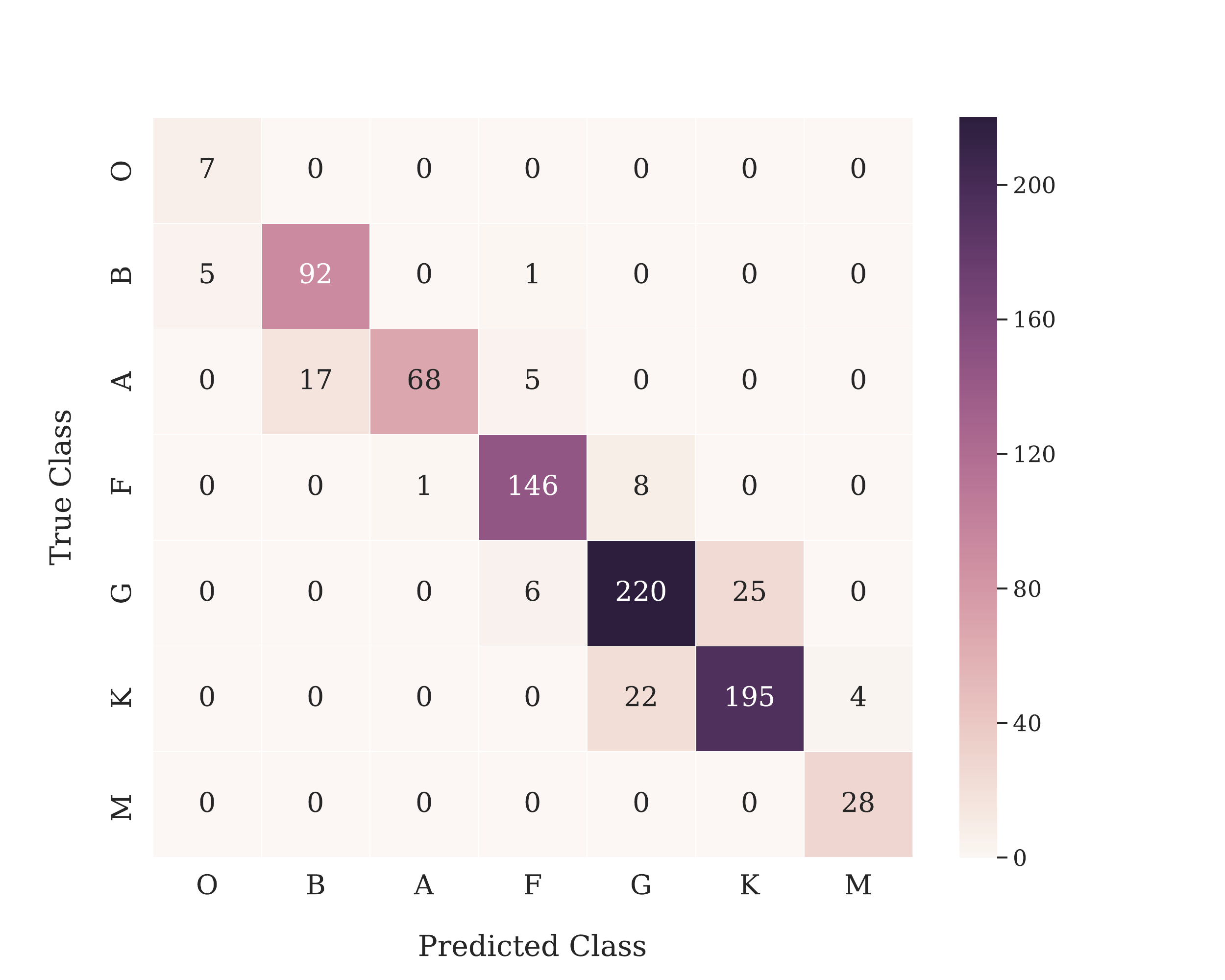}
\caption{Confusion matrix for the main spectral classes computed for the CFLIB library using 1-D CNN classification model.}
\label{fig:seta_cm_ae}
\end{figure}

Once the autoencoder model parameters are adjusted, we design a classifier model using the encoding structure of the autoencoder model and adding flattening layers to convert multi-dimensional data to one-dimensional form. The last layer classifies the input features into one of the 70 subclasses. This classification model is trained on our training set and testing is performed on the CLFIB library following the same procedure as described in Sec.~\ref{sec:ml_methods}. Comparison of the predicted classes with respect to the expected classes is shown in Fig.~\ref{fig:ae_seta_classification} and the class-wise confusion matrix is shown in Fig.~\ref{fig:seta_cm_ae}.

\begin{figure}
\centering
\includegraphics[scale=0.28]{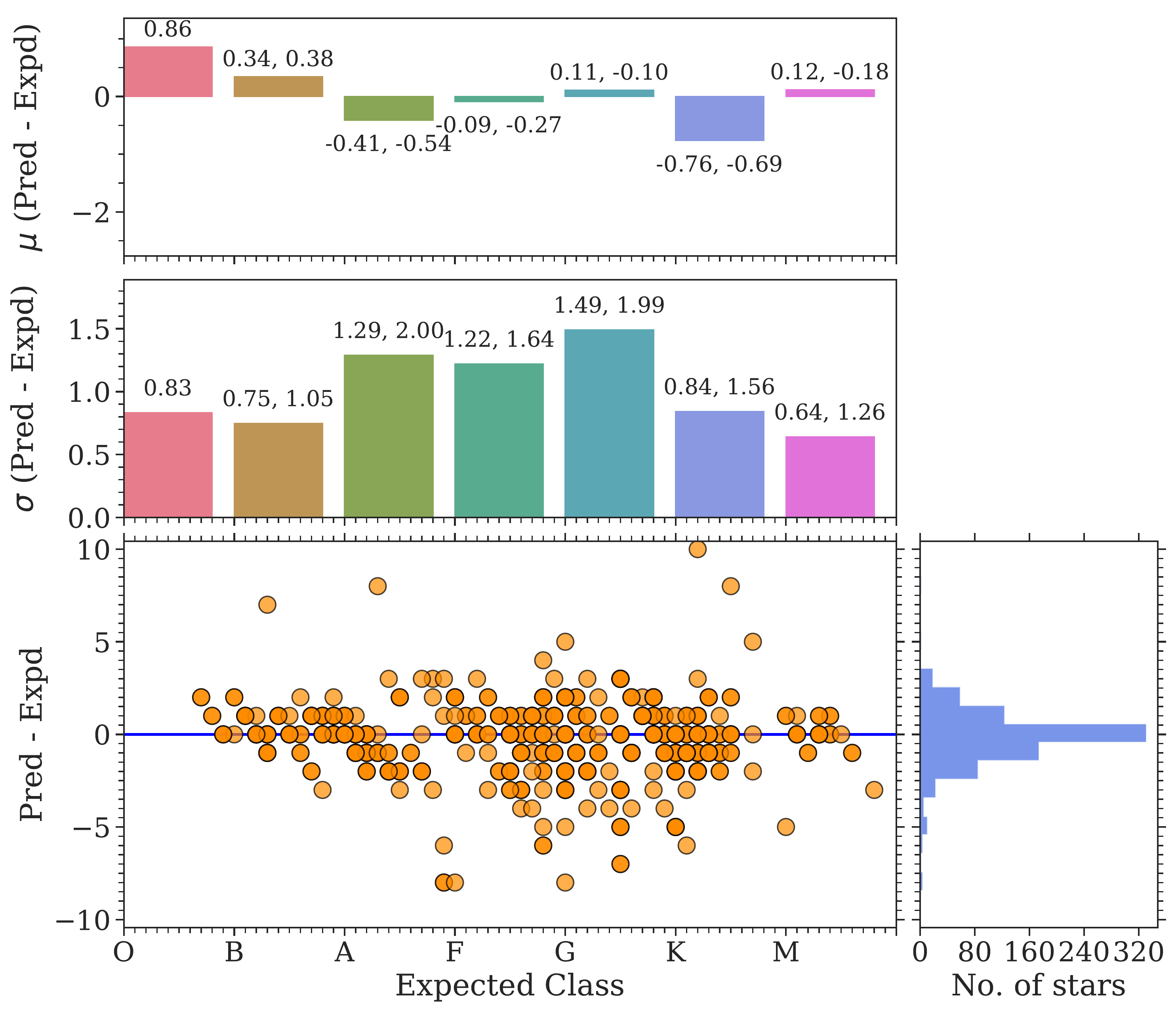}
\caption{Classwise error statistic. Scatter plot in the left bottom panel shows the classification errors with respect to spectral classes with a histogram of the errors at the right. Bar-plots in the top and middle panels show the mean and standard deviations for each main stellar class. Each mean and standard deviation statistic has two numbers, one computed with outliers removed with 3$\sigma$ clipping followed by the other which has been computed without removing any outliers.}
\label{fig:seta_ae_errplt}
\end{figure}

Excluding 3$\sigma$ outliers, we obtain a mean difference of $-0.14$ between the predicted and expected classes with a standard deviation of $1.23$ subclasses. The implied error of 1.23 subclasses achieved using autoencoders is slightly better than the error of 1.35 subclasses obtained using a shallow neural net. Difference between the predicted and expected class for each class is illustrated in Fig.~\ref{fig:seta_ae_errplt}, where the mean and standard deviation of the difference is shown for each main class as a bar plot. In those cases where normal statistics (mean and standard deviation) are different from sigma-clipped statistics (3$\sigma$), we provide both estimates, with the sigma-clipped estimate followed by the normal estimate, and put a single value where both have the same value. For example, A type spectra have a mean difference of $-0.54\pm2.00$ whereas 3$\sigma$ clipping gives a mean difference of $-0.41\pm1.29$. The top panel shows the mean difference $\mu\,(\textrm{Pred}-\textrm{Expd})$ which indicates that all classes are biased within 1-subclass only.

For each individual class, the precision, recall and F1-score are presented in Table~\ref{tab:seta_report_ae}.

\begin{table}
\caption{Classification report for the main classes using Convolutional Autoencoder.}
\centering
\begin{tabular}{lcccc}
\hline\hline
Class & Precision & Recall  & F1-score & Population \\\hline
     O & 1.00      & 1.00    & 1.00     & 7         \\
     B & 1.00      & 0.88    & 0.93     & 98        \\
     A & 0.87      & 0.93    & 0.90     & 90        \\
     F & 0.86      & 0.92    & 0.88     & 155       \\
     G & 0.82      & 0.89    & 0.85     & 251       \\
     K & 0.94      & 0.82    & 0.88     & 221       \\
     M & 0.90      & 0.96    & 0.93     & 28        \\
 Total & 0.89      & 0.88    & 0.88     & 850       \\
\hline
\end{tabular}\label{tab:seta_report_ae}
\end{table}

\subsection{Luminosity Classification: Method and Results}\label{sec:lum_classification}

For classifying stellar spectra into the luminosity classes I, II, III, IV, V, we explored shallow as well as deep learning architectures and find that 1-D convolutional neural networks (1-D CNN) with autoencoders give the best results. For constructing a luminosity classifier, we adopt the same approach as in the case of spectral classification. The encoder part of the autoencoder architecture (Fig.~\ref{fig:ae_architecture}, Sec.~\ref{sec:deep_net}) remains the same. Encoding layers are appended with a few fully connected layers which finally converge to the output layer with 5 nodes. The full network architecture is as follows:

\begin{Verbatim}[fontsize=\small]
# ENCODER
input_spec = Input(batch_shape=(None,2896,1))
encoded = Conv1D(128, 3, activation=`relu',
                 padding=`same')(input_spec)
encoded = MaxPooling1D(2)(encoded)
encoded = Conv1D(64, 3, activation=`relu',
                 padding=`same')(encoded)
encoded = MaxPooling1D(2)(encoded)
encoded = Conv1D(32, 3, activation=`relu',
                 padding=`same')(encoded)
encoded = MaxPooling1D(2)(encoded)
encoded = Conv1D(16, 3, activation=`relu',
                 padding=`same')(encoded)
encoded = MaxPooling1D(2)(encoded)

# CLASSIFIER
clf_layer1 = Dense(64, activation=`softsign')(encoded)
clf_layer2 = Dense(32, activation=`softsign')(clf_layer1)
clf_layer3 = Dense(16, activation=`softsign')(clf_layer2)
clf_layer4 = Dense(5, activation=`softmax')(clf_layer3)
\end{Verbatim}

The number of neurons in the classification layers are chosen after experimenting with different combinations. For training the above network, we load the encoder model with the adjusted weights obtained during the unsupervised pre-training phase on SDSS data. To determine the classification layer weights for the supervised classification, we retrain the network with the training set as an input and the corresponding luminosity classes as the output. For the training, we perform a 85-15\% training and validation split and monitor the categorical cross-entropy (Eq.~\ref{eq:loss_function}) loss function for early-stopping the training. Once the loss-function for the validation set is minimum and does not improve for consecutive epochs, the training stops. This is done in order to avoid over-fitting. After the final epoch of training, we get an accuracy of 84\% and 77\% for the training and validation set respectively. 

\begin{figure}
\centering
\includegraphics[scale=0.39]{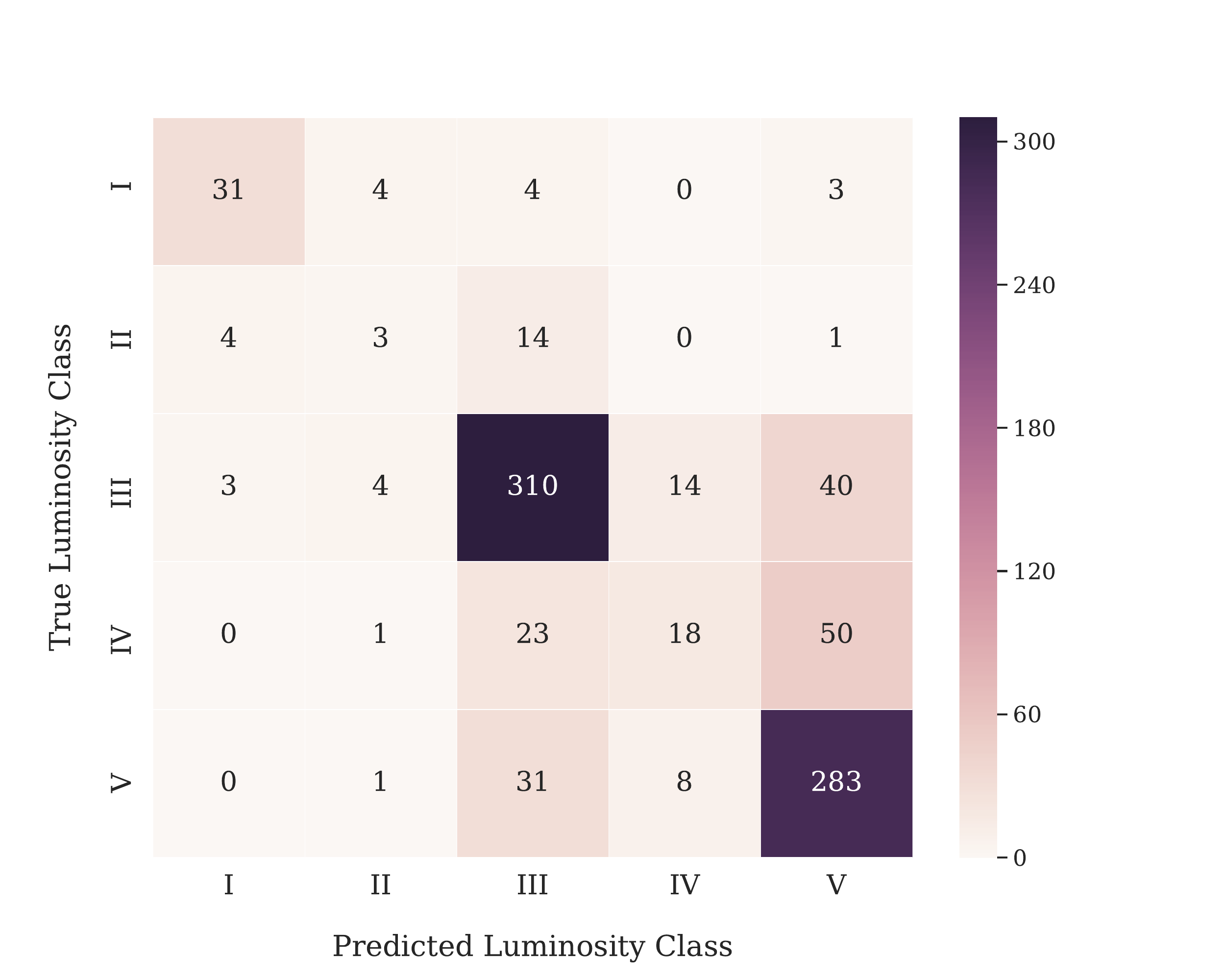}
\caption{Confusion matrix for luminosity classification on CFLIB using 1-D Convolutional Neural Network (1-D CNN).}
\label{fig:lumcls_cm_ae}
\end{figure}

Finally, the trained network is tested on the 850 spectra from CFLIB. Comparing the predicted luminosity classes with the true labels from the literature gives an overall accuracy of 76\%. The corresponding confusion matrix is shown in Fig.~\ref{fig:lumcls_cm_ae}.

To obtain the error in the luminosity classification, we label the classes I-V with numbers 0-4 respectively. We compute the mean and standard deviation between the predicted and the true class (i.e. $\textrm{code}_{\textrm{pred}}-\textrm{code}_{\textrm{true}}$) after removing 3$\sigma$ outliers. The mean value comes out to be 0.09 with standard deviation of 0.76 (mean and sigma of 0.07 and 0.72 respectively after removing outliers). Spectral and luminosity classes for the 850 stars selected from CFLIB are presented in Table~\ref{tab:cflib_pred}.

\begin{table}
\caption{Expected and predicted class for CFLIB spectra. ``C Flag'' in the last column indicates the consistency between the two spectral classifications. ``0'' indicates a good agreement between the expected and predicted classes (within 3$\sigma$) whereas ``1'' indicates that the classifications are inconsistent by more than five subclasses. The most discrepant cases are discussed in detail in Sec.~\ref{sec:misclassification}. The table is published in its entirety in the electronic version. A portion is shown here for guidance regarding its form and content.}
\centering
\begin{tabular}{lccc}
\hline\hline
Identifier      & Expected Class & Predicted Class & C Flag \\\hline
HD102212        & M1 III    & M1 III   & 0 \\
HD10307         & G1 V      & G0 V     & 0 \\
HD10476         & K1 V      & K2 V     & 0 \\
HD118098        & A3 V      & A2 V     & 0 \\
HD121370        & G0 IV     & G0 IV    & 0 \\
HD149757        & O9 V      & O9 V     & 0 \\
HD164136        & F2 II     & F2 II    & 0 \\
HD164353        & B5 I      & B5 I     & 0 \\
BD+45 1668      & A9 III    & A1 V     & 1 \\
G 63-9          & F9 V      & F8 V     & 0 \\
\hline
\end{tabular}\label{tab:cflib_pred}
\end{table}

We note that there is a well-known correlation between the temperature and luminosity classes. Massive hot stars evolve faster and hence likely to find them during their main sequence phase. In the case of a magnitude limited sample covering G-M stars, more number of giants will be targeted relative to the faint dwarfs. Combination of the two effects gives a large number of OBA dwarfs and GKM giants, hence a correlation between spectral type and luminosity class can be seen in a magnitude limited sample. In our test set, CFLIB, the ratio of dwarfs to giants is about 3 for hotter stars. For cooler spectra, the ratio of giants to dwarfs is about 2.5, which shows that OBA type spectra are dominated by dwarfs and GKM types are dominated by giants. In such a situation, a luminosity classifier which assigns luminosity classes only based on the temperature classes, i.e., classifies all hotter stars as dwarfs and cooler stars as giants, without any reference to the features in the spectra will also give a reasonably good performance. For such shallow and inefficient classifier, we find that the values of mean difference and standard deviation would be close to 0.20, 1.2 classes respectively for our test sample. To ascertain that the classification strategy adopted by our luminosity classifier indeed relies on the spectral features and is not governed by the skewed population of dwarfs and giants, we perform the following test. We consider all G type spectra in our test sample, which contains roughly equal population of dwarfs and giants (123 belonging to class III and 88 belonging to class V). If the classifier is influenced by the temperature classes and does not learn the feature-based classification, we would expect the majority of G type spectra being classified as giants, which would lead to a poor classification of this sub-sample with a larger RMS error.

We apply our trained luminosity classifier on this sample and get an accuracy of about 80\% with $F1$ score of 0.85 and 0.86 for class III and V respectively. We obtain an RMS error of 0.62 classes (0.36 after removing 3$\sigma$ outliers) for this sub-sample. We also repeat the same exercise for B type spectra and obtain an RMS error of 0.8 classes. This analysis approves the feature-based classification by our model rather than just assigning the luminosity classes based on temperature classes.

\section{Outliers and Classifier sensitivity}\label{sec:misclassification}

In Table~\ref{tab:classification_results}, we have shown that using a 1-D CNN model with autoencoder provides the least 1$\sigma$ error of 1.23 subclasses in the spectral classification. Fig.~\ref{fig:ae_seta_classification} shows the comparison between the expected and predicted classes. We note that there are a few spectra for which the predicted class disagrees with the expected class substantially. In this section, we study each outlier to understand the origin of the discrepancy. We also perform some limitation checks to evaluate the degradation in the performance of the model when the input spectra are in the continuum-normalized format or suffer from incorrect/incomplete data reduction processes.

\subsection{Discussion on Outliers}\label{sec:outliers}

\begin{figure*}
\centering
\includegraphics[scale=0.32]{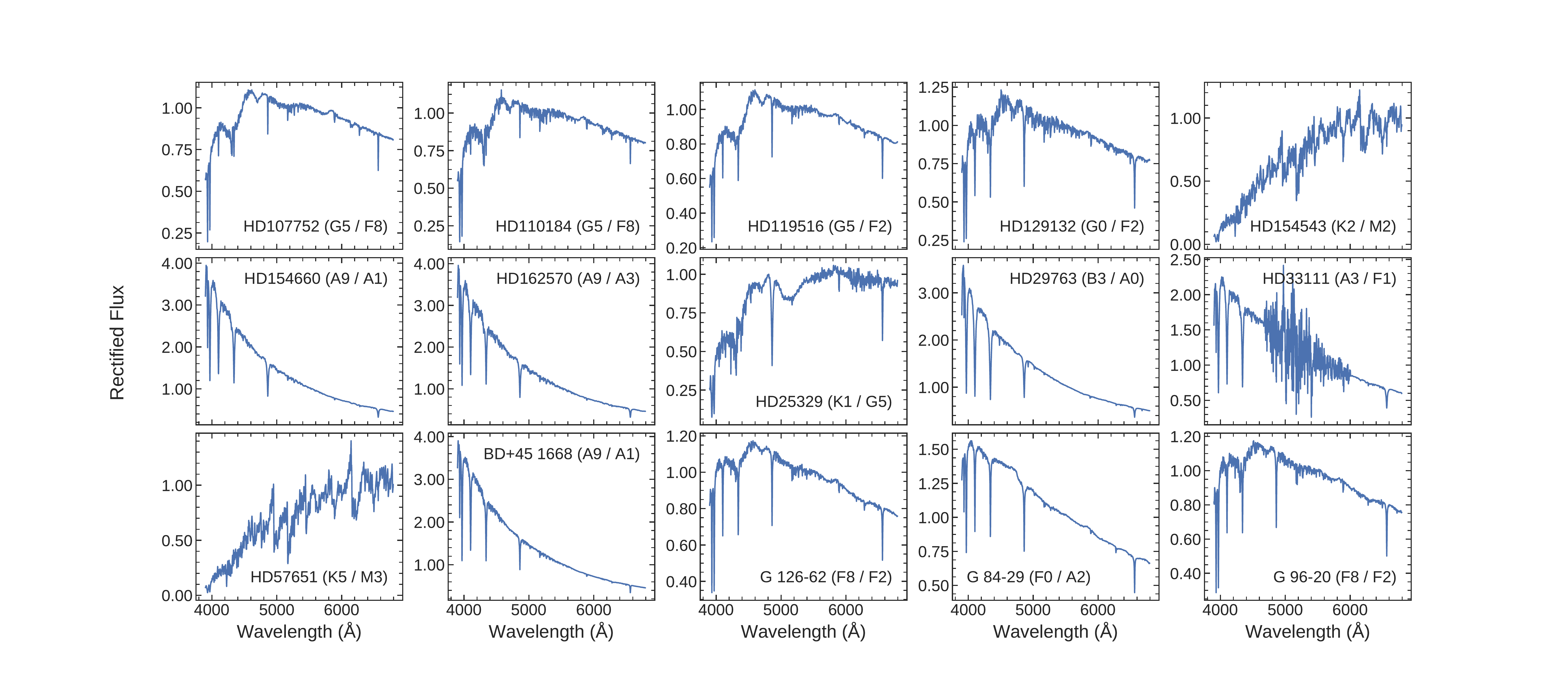}
\caption{CFLIB spectra which are misclassified by more than five subclasses using CNN with autoencoder. Each panel shows pre-processed spectrum corresponding to misclassified object along with the object identifier. Expected and predicted classes are provided in the parentheses separated by a slash (`/') symbol.}
\label{fig:misclassified_ae}
\end{figure*}

We investigate those cases in detail which have been misclassified by our model by more than five subclasses (half main class). Spectra of these stars are presented in Fig~\ref{fig:misclassified_ae} and their details are presented in Table~\ref{tab:misclassified_cflib}.

\begin{table*}
\caption{Details of misclassified CFLIB spectra. The atmospheric parameters are adopted from \citet{Wu2011}.}
\centering
\begin{tabular}{lccrrrl}
\hline\hline
Identifier   & Expected & Predicted & \multicolumn{3}{c}{Parameters \citep{Wu2011}} & Remarks \\
             &  Class   &   Class   &     \teff{}    &     \logg{}  &     \feh {}   &         \\\hline
HD107752     &   G5     &    F8     &    4868        &    1.90      &    $-2.47$    & Metal-poor star; larger relative sensitivity for flux error \\
HD110184     &   G5     &    F8     &    4478        &    0.96      &    $-2.25$    & Metal-poor star; larger relative sensitivity for flux error \\
HD119516     &   G5     &    F2     &    5315        &    2.62      &    $-1.84$    & Metal-poor star; flux calibration issue \\
HD129132     &   G0     &    F2     &    6761        &    3.90      &      0.04     & Predicted class consistent with the atmospheric parameters \\
HD154543     &   K2     &    M2     &    3576        &    1.27      &    $-0.03$    & Predicted class consistent with the atmospheric parameters \\
HD154660     &   A9     &    A1     &    7663        &    3.97      &    $-0.18$    & Tentative classification \\
HD162570     &   A9     &    A3     &    7511        &    3.87      &      0.02     & Tentative classification \\
HD25329      &   K1     &    G5     &    4840        &    4.85      &    $-1.68$    & Flux calibration issue; metal-poor star \\
HD29763      &   B3     &    A0     &   10073        &    2.60      &    $-0.51$    & Predicted class consistent with the atmospheric parameters \\
HD33111      &   A3     &    F1     &    8002        &    3.78      &    $-0.20$    & Flux issue \\
HD57651      &   K5     &    M3     &    3418        &    0.85      &    $-0.07$    & Predicted class consistent with the atmospheric parameters \\
BD+45 1668   &   A9     &    A1     &    8693        &    4.72      &    $-0.48$    & Flux calibration issue \\
G 126-62     &   F8     &    F2     &    6055 	     &    4.09      &    $-1.44$    & Metal-poor star; larger relative sensitivity for flux error \\
G 84-29      &   F0     &    A2     &    6189        &    3.88      &    $-2.25$    & Metal-poor star; larger relative sensitivity for flux error \\
G 96-20      &   F8     &    F2     &    6284        &    4.26      &    $-0.90$    & Consistent with other literature source; tentative classification\\
\hline
\end{tabular}\label{tab:misclassified_cflib}
\end{table*}

\begin{itemize}

\item[--] HD107752: This is a high proper motion and metal-poor star with \feh{}\,=\,$-2.74$ \citep{Beers2000}. Expected class for this star is G5 whereas our model assigns a hotter class F8. In a high-resolution spectroscopic study ($R\sim 22,000$), \citet{Burris2000} find its atmospheric parameters: [\teff, \logg, \feh]\,=\,[4700, 1.70, $-2.69$] (in the rest of the paper we adopt this bracketed notation for parameters). The atmospheric parameters support its later G or early K-type spectral class which makes our classification incorrect. We believe that the assignment to a hotter class is due to its metal-poor nature which are not well represented in our training sample.

\item[--] HD110184: Classified as G5 \citet{Cannon1993} in our literature compilation, this star is also assigned a hotter temperature class, F8, by our model. \citet{Mishenina2001} analyse its high-resolution spectrum and obtain \teff{}\,=\,4380\,K, \logg{}\,=\,0.60\,dex with \feh{}\,=\,$-2.27$\,dex which implies its metal-poor nature. We do not find any discrepancy in its CFLIB spectrum. Lower metallicity is the most probable reason which makes the classifier to predict it as a hotter star.

\item[--] HD119516: Expected class for this star is G5 while our model predicts it as F2 type. Effective temperature of 5689\,K with \logg{}\,=\,2.23\,dex in \citet{Behr2003} makes it more consistent with a late F spectral type, e.g. F8. These parameters are estimated based on detailed spectrum synthesis, line profile analysis and ionisation equilibrium using high resolution echelle spectra (R\,=\,60,000). Our classification is still hotter by six subclasses as compared to revised spectral class F8. \citet{Behr2003} also estimates \feh{}\,=\,$-1.92\pm0.05$\,dex. We find that this is another metal-poor star which is being assigned a hotter class by our classifier due to lack of representation of such stars in the training.

\item[--] HD129132: The spectrum is classified as G0V \citep{Harlan1970}. The model predicts it to be an F2V type star. \citet{Wu2011} estimate the atmospheric parameters for this object using low-resolution spectrum and find [6761, 3.90, 0.04]. A high resolution (R\,=\,60000) spectroscopic abundance study by \citet{Luck2017} finds \teff{}\,=\,6609\,K and \logg{}\,=\,3.41\,dex. Estimated \teff{} values from high and low-resolution spectroscopic studies seem to be more consistent with our predicted class.

\item[--] HD154543: Expected class for this star is K2I whereas our model predicts it to be an M1II type. Not much information is available in literature for this star. Atmospheric parameters in \citet{Wu2011} are [3576, 1.27, $-0.03$] which seem to agree with our classification. Therefore, we believe that our classification is reliable.

\item[--] HD154660, HD162570: In the classification scheme of CFLIB, these two spectra have been classified as A9V whereas our model predicts hotter subclasses A1V and A3V respectively. \citet{Wu2011} estimated their parameters as [7663, 3.97, $-0.18$] and [7511, 3.87, 0.02] respectively. In the Michigan catalog, \citet{Houk1999} classified HD154660 as an A5V type star, while \citet{Paunzen2001} classify it as A3III using template matching based on low resolution spectra covering a narrow wavelength region around H$\gamma$. Unfortunately, there are no high resolution spectroscopic studies available for these objects. Our results can be considered tentative.

\item[--] HD25329: This high-proper motion K1 type star has been classified as G5 by our model. \citet{Gray2003} classify it as K3 type with [4889, 4.83, $-1.61$]. \citet{Wu2011} also determine similar parameters: [4840, 4.85, $-1.68$]. \citet{Sharma2016} estimated similar values of [4964, 4.60, $-1.58$] for these parameters using the full spectrum fitting technique. Spectral type vs \teff{} calibrations suggest that the K1 class is consistent with \teff{} $\sim$ 5000\,K. There are issues with the flux calibration of this spectrum as shown in the figure. We believe that poor flux-calibration along with its metal-poor nature are the main source of misclassification.

\item[--] HD29763: This is a spectroscopic binary system where the primary component is assigned B3V class \citep{Lesh1968}. Using our classification model, we assign A0V class to its spectrum. \citet{Hamdy1993} assign B5 class to this star based on photometric calibrations. The estimated parameters for this star are [10073, 2.60, $-0.51$] from \citet{Wu2011} which signals that the star is of late B/early A type and agrees with our assigned class.

\item[--] HD33111: For this star, our model predicts F1 class which is cooler than the assigned class in the literature A3 \citep{Gray2003}. The atmospheric parameters [8002, 3.78, $-0.20$] from \citet{Wu2011} are in more agreement with the literature class. We notice that the CFLIB spectrum for this star, as shown in Fig.~\ref{fig:misclassified_ae}, is very noisy in the mid-optical region which could be a potential reason for the misclassification. To confirm this, we remove the noisy features, replace that region by the respective continuum and redo the classification. The updated class comes out to be A2 which is in agreement with the original class.

\item[--] HD57651: In CFLIB, the assigned class to this spectrum is K5 but it is classified as M3 using our classification model. The estimated \teff{} and \logg{} values for this spectrum are 3418 K and 0.85 dex respectively, which makes the M3 class more plausible than K5, which would correspond to a hotter temperature of $\sim$ 4200\,K.

\item[--] BD+45 1668: This is a RR Lyr star of A9 type according to literature sources. Our model assigns it to class A1. Estimated parameters for this spectrum in \citet{Wu2011} are [8693, 4.72, $-0.48$], while some other works estimate a lower temperature, close to 6300 K \citep{Liu1990,Sandage1993,Kovacs2003}. We believe that our misclassification is because of the SED which matches with that of a mid or late B type star.
  
\item[--] G 126-62: It is another metal-poor star of our test sample for which the assigned class, F2, is hotter by 6 subclasses as compared to the compiled class, F8. \citet{Gratton2003} estimated its atmospheric parameters as [6085, 4.12, $-1.58$] using high-resolution spectroscopic analysis. It is also a member of spectroscopic binary system \citep{Pourbaix2004} but the spectral type for the secondary component is not known. \citet{Gorgas1999} also classify it to F8 class with \feh{} value of $-1.90$. \citet{Bidelman1985} assigns class F2 to its spectrum based on visual inspection. Class F8 agrees more with the estimated \teff{} value for this object and the reason for discrepancy could be its metal-poor nature.

\item[--] G 84-29: This star is very well studied in the literature because of its metal-poor nature. The expected class for this halo star \citep{Hobbs1988} is F0V whereas the model classifies it as A2IV type. \citet{MacConnell1971} classify it as A type horizontal branch star by studying Balmer lines its spectrum. \citet{Lee1984} classify this star as sdF0 (subdwarf of F0 type) by analyzing its low-resolution spectrum. Estimated atmospheric parameters for this object are [6189, 3.88, $-2.25$] from \citet{Wu2011} whereas \citet{Casagrande2010} determine a hotter \teff{} with the parameters [6419, 3.97, -2.70]. \citet{Mashonkina2008} study this metal-poor star extensively to derive the three parameters and detailed abundances using its high-resolution, high SNR ($\sim$100) spectrum. They obtain [6340$\pm100$, 3.90$\pm0.15$, $-2.65\pm$0.01]. As discussed in \citet{Fuhrmann1998}, this star suffers from interstellar reddening. The explanation previously provided for metal-poor stars HD107752, HD110184, and G 126-62 holds true for this object as well and might be the reason for misclassification.

\item[--] G 96-20: This high-proper motion star is classified as F8 in our compilation for which \citet{Nissen2014} estimate [6445, 4.46, $-0.90$] from high-resolution spectroscopic study. \citet{Nesterov1995} classify it as F5 based on visual examination. Our model assigns a hotter class F2 which is not very discrepant with respect to the spectral class from \citet{Nesterov1995}. We conclude that our classification is tentative for this source.

\end{itemize}

We find that in some cases discussed cases, our classification is consistent with the classification/atmospheric parameters provided in the literature. In a few cases, we provide tentative classification as we could not trace the exact origin for the misclassification but in these cases, our model assigned classes seem to be consistent by visual inspection. There are a few cases where the misclassification is caused either due to the wrong flux calibration of CFLIB spectra or due to the extremely metal-poor nature of the objects. We also realize that our training set contains only six metal-poor stars with \feh{}\,<\,$-2.0$\,dex, which might be the factor responsible for misidentification in the lower metallicity regime which requires a good representation of spectral features in the spectra. In the metal-poor regime, it seems that because of poor lines in the spectra, the relative sensitivity for flux error is getting larger which might be causing the error in the classification.

\subsection{Classifier Sensitivity}\label{sec:sensitivity}

\begin{figure*}
\centering
\includegraphics[scale=0.45]{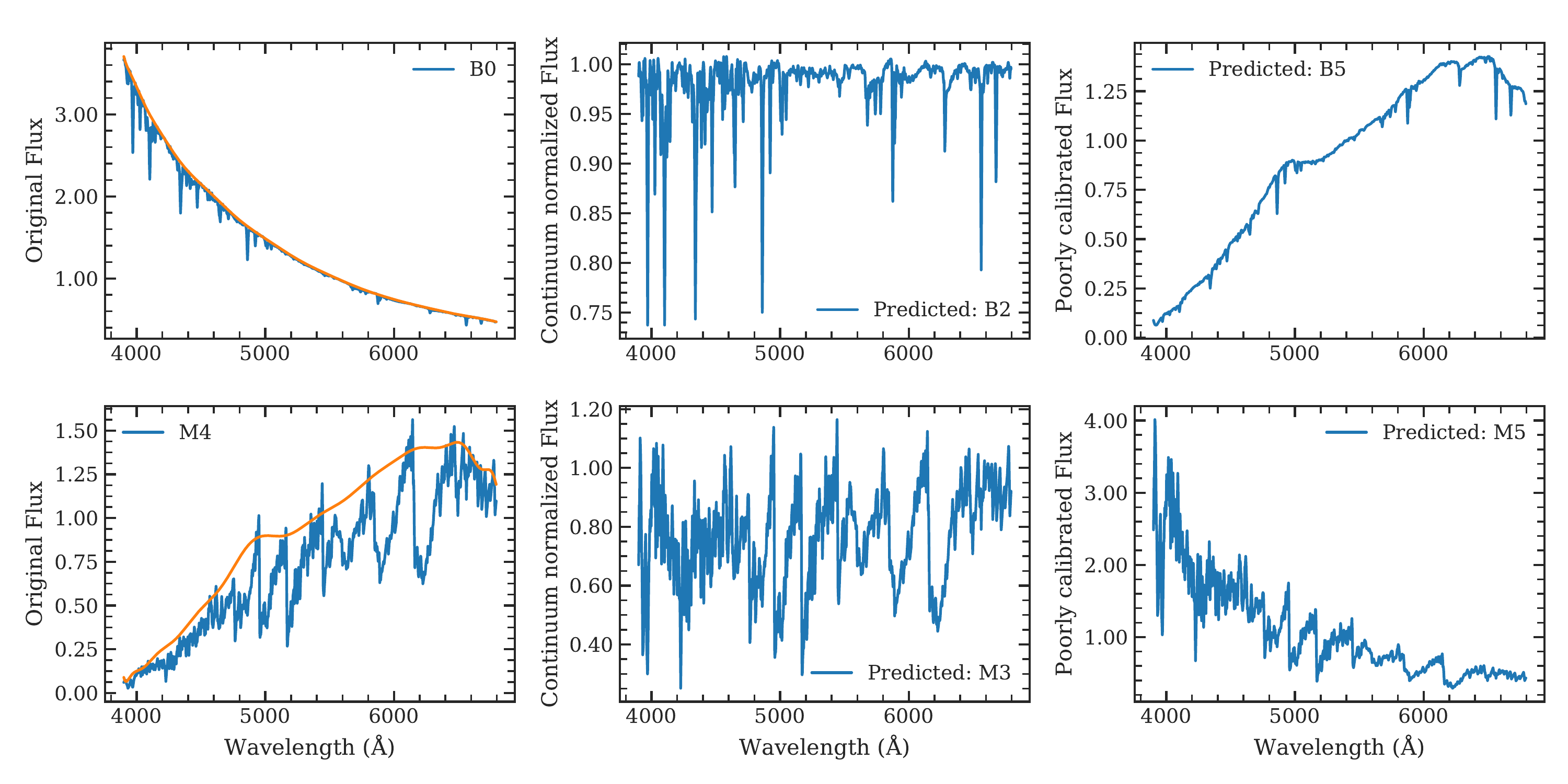}
\caption{Two sample CFLIB spectra for HD184915 of type B0 (upper panels) and HD175588 of type M4 (lower panels). The left panels show the original pre-processed spectra (blue) with the approximated polynomial representing the continuum level (orange). The middle panels show the spectra where the continuum is normalized to unity. In the two right most panels, the continuum of each spectrum is replaced by the continuum of other star (see point 3 in Sec.~\ref{sec:sensitivity}). The predicted classes by the CNN model for normalized spectra are indicated in the legend.}
\label{fig:flux_issues}
\end{figure*}

For investigating the sensitivity of our classifier on various issues related to the spectral energy distribution (SED), the robustness of the model with respect to wrong labels in the training set, errors in the flux-calibration or continuum, and interstellar reddening, we perform some diagnostics which are discussed below:

\begin{enumerate}[label=\arabic*., listparindent=0.5em]

\item Continuum Normalization: The classification models presented in Sec.~\ref{sec:techniques} are trained and tested on the flux-calibrated spectra, i.e. the observed counts for a source are converted to actual flux values using spectra of standard stars. The major reason for us to develop classifiers for flux-calibrated spectra is the fact that large-scale surveys which can benefit from such models provide spectra in flux-calibrated format. However, in some cases, an individual user may have spectra only in the continuum-normalized format where the spectral continuum is normalized to unity everywhere and only spectral features are left in the spectrum. To check the performance of our classifier on such continuum normalized spectra, we normalize our test sample of 850 spectra by applying the Savitzky-Golay filter \citep{Savitzky_Golay1964} of different sizes and fitting a polynomial to the filtered and smoothed flux. Examples of continuum normalized spectra obtained through this method are shown in the middle panels of Fig.~\ref{fig:flux_issues} for one B0 one M4 type.

We first apply the Keras ANN classifier (trained on flux-calibrated spectra) to this sample and get a RMS error of 4 subclasses with a systematic difference of $-0.2$ subclasses. With the deep CNN model, we obtain an improved error of 3.04 subclasses and roughly the same value of the systematic difference between the expected and predicted classes. This shows that while the performance degrades if the continuum is not present in the test spectra, it does not cause a systematic shift of spectra from one class being assigned to another quite different class by a huge difference. The errors do not shoot up, for example, to 10 subclasses which would mean misclassification by one complete temperature class. This also provides evidence that our model does not simplify the classification problem by just relying on the continuum but does indeed learn the features present in the spectra. We also carried out another test where we normalize our training sample in the same manner as described above and train the ANN architecture (Fig.~\ref{fig:seta_nn_classifier}). We test the continuum-normalized ANN model on normalized CFLIB spectra and obtain an RMS error of 1.81 subclasses, which is higher than the error of 1.35 subclasses obtained using ANN on flux-calibrated spectra.

We note that the continuum normalization is a tedious task which requires much attention \citep[e.g.][]{Fabbro2018}. A simple polynomial fitting routine might not be adequate for normalizing the continuum for a range of spectral types and varied S/N spectra. Continuum normalizing is challenging especially for cool stars, carbon stars, stars with strong spectral features and emission lines. It is also a quite subjective process from person to person. However, this approach gives us fairly good representation of normalized spectra to get an estimate of classification accuracy in continuum-normalized cases.

\item Wrong classifications in training set: To prepare a larger sample for training the deep convolutional neural network, we have no option other than combining spectra from different databases. Spectra in these source libraries have been classified using different approaches by different observers and therefore might suffer from systematic differences between the different classification methodologies. Although we try our best to make sure that none of the spectra are supplied a wrong class by querying each identifier in SIMBAD as well and examining the spectra in all the subclasses for any erroneous spectra entering our training set as described in Sec.~\ref{sec:pre-processing}. However, there is a non-zero probability of a very few spectra being assigned wrong labels. We argue that given the choice of categorical cross-entropy as the model loss function (Eq.~\ref{eq:loss_function}), the training would be robust to outliers as the outliers will be heavily penalized by the softmax activation function in $q(x)$.

To verify it, we randomly shuffle classes assigned to 40 spectra (about 2\% of the sample) in the training set, making sure that these 40 examples are spread over all the temperature classes. We retrain the Keras ANN classifier with this impure training set and test the trained model on the CFLIB spectra. We observe that the RMS error goes to 1.37 subclasses as compared to 1.35 subclasses obtained with the clean sample. We then increase the number of incorrect classes to 100 ($\sim$\,5\%) and find that the error goes up marginally to 1.46 subclasses. This test supports our claim regarding the robustness of the method against a small fraction of incorrect labels. Repeating the same exercise with 200 spectra ($\sim$\,10\%) spectra in the training set increases the error to 1.8 subclasses with an increased number of 3$\sigma$ outliers.

\item Improper flux-calibration: As listed in Table~\ref{tab:misclassified_cflib}, there are a few CFLIB spectra which are not correctly labelled by our classifier due to incorrect flux-calibration, e.g. HD25329. To analyze the effect of incorrect flux-calibration we implement another test. We obtain the continuum for each spectrum in our test set as described previously, shuffle the continuum randomly, and multiply each shuffled continuum by the normalized spectrum it is assigned, to generate a sample in which all spectra have incorrect flux-calibration. Two example spectra produced by this exercise are presented in the right panels of Fig.~\ref{fig:flux_issues}. We classify this sample using our best model and find the RMS error of 4.5 subclasses between the expected and predicted classes. We repeat this exercise 10 times to generate 10 test sets with flawed flux-calibration and obtain the RMS error lying between 3.5-6.0 subclasses. As expected, there is a significant decline in the classifier performance but the classifier can still recover the actual class with an upper bound of 6.0 subclasses RMS error in a worst-case scenario.

\item Interstellar reddening: The absorption of the incoming stellar photons by the interstellar medium can significantly affect the observed spectrum and therefore should be corrected for the extinction in accordance to the extinction laws and the amount of extinction in the direction of the target. However, if there are no available measurements for the interstellar absorption in the line-of-sight direction, how accurate our classifier would be? To check this aspect, we consider all 850 CFLIB spectra and artificially introduce reddening with different visual absorption values ($A_V$) in the range from 0.0 to 1.0 in steps of 0.01. Reddening is applied using \citet{Fitzpatrick2007} extinction model for $R_v$ value of 3.1. We find out that the as we keep on increasing the value of reddening, the performance degrades to RMS error of 1.52 subclasses at $A_V=0.1$, to 3.48 subclasses at $A_V=0.3$. At $A_V$ equal to 0.4 the error goes to 5.09 subclasses which remains almost steady till $A_V$ value of 0.80. At $A_V=1.0$, the error increases to 11.87 subclasses.

To check how an expert classifier would perform on this sample, we provide the same reddened spectra to MKCLASS classifier \citep{Gray2014} and use both the libraries, `libnor36' and `libr18', to classify the spectra in automated mode using a script. The best results are obtained with libr18 which gives an RMS error of 2.04 subclasses at $A_V\,=\,0.0$ and is able to assign the spectral class for 753 spectra, though in a few cases, luminosity class is missing. Remaining spectra are either not assigned any class (labelled by a ``?'': 57 cases) or assigned a class with peculiarity (e.g. `kA3hF1mF2 Eu', 40 cases ). For the spectra with $A_V$ values of 0.5 and 0.6, the error goes up to 7.2 subclasses and 8.0 subclasses respectively. Increasing the reddening further does not cause any significant change in the error but the number of classified spectra goes down to about 550 at $A_V\,=\,1.0$ with RMS error of 7.8 subclasses.

\end{enumerate}

We conclude that our classification model can tolerate the defects in the stellar spectrum with a limited decline in the classification accuracy. For the normalized spectra, the classification accuracy is 3.1 subclasses whereas the spectra with improper flux-calibration can result in a classification with an upper limit of RMS error of 5.1 subclasses. We notice that interstellar reddening plays an important role in assigning a correct class and our classifier can tolerate the visual absorption up to the value of 0.4 with an error of 5.1 subclasses, after which there is a sharp decline in the performance. We observe a similar decline in MKCLASS classifier as well which shows that the incorrect reddening estimate will equally hamper the performance of a human(like) classifier. This is especially true for spectra with lines which are widely separated in wavelengths and are not continuum normalised. Therefore, it is advisable to correct the spectra for reddening before applying any automatic classification model. Extinction correction becomes fairly straightforward with the availability of extinction maps by \citet{Schlegel1998}, \citet{Schlafly2011}, etc.

\section{Application to SDSS}\label{sec:sdss_application}

We apply the 1-D CNN classifier trained as described in Sec.~\ref{sec:deep_net} to a sample of stellar spectra from the SDSS DR13 database. All spectra in the SDSS database are classified into three primary classes: `STAR', `GALAXY' or `QSO'. Stellar spectra are further classified into subclasses using the SDSS pipeline \texttt{spec1d} \citep[See Sec. 4,][]{Bolton2012} which provides the spectral class as well as the luminosity class (though luminosity classes are not available for all spectra). For each spectrum, the pipeline also estimates the median signal-to-noise ratio. We use this information to query the SDSS DR13 database and choose all objects which are classified as `STAR', have SNR greater than 20 and for which the wavelength coverage is complete (3900-9500 \AA). Our SQL query returns a table with 61,627 unique spectroscopic observations with their SDSS object identifiers such as object ID, plate ID, MJD, fiber ID and celestial coordinates along with the primary class (STAR/GALAXY/QSO), subclass (MK Class) and SNR assigned by the SDSS pipeline. Using plate ID, MJD and fiber ID, we download the 61,627 stellar spectra. Only 48,084 of these spectra have complete MK classification available while the remaining spectra either have missing spectral and/or luminosity class or have been classified to classes like carbon stars, WD (white dwarfs), Calcium WD, CV etc. We supply the SDSS spectra with complete MK classification to our trained classifier after pre-processing as described in Sec.~\ref{sec:pre-processing}. 

\begin{figure*}
\centering
\includegraphics[scale=0.4]{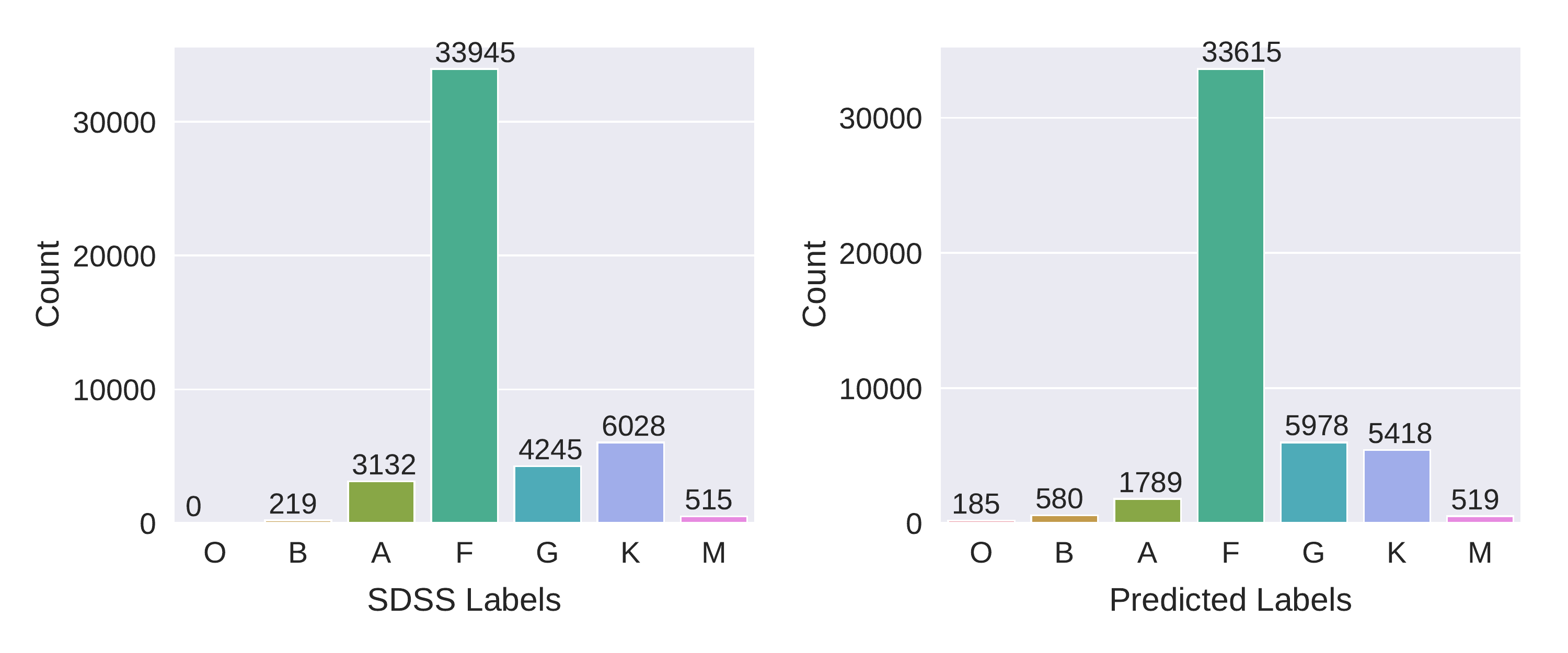}
\caption{Main spectral class distribution for the SDSS sample. The left panel shows the distribution according to the classification provided by SDSS whereas right panel uses our model predicted classes.}
\label{fig:sdss_all}
\end{figure*}

To investigate the CNN based classifier at the main class and sub-class level, we first check the distribution at the main class level according to the SDSS pipeline and our model. As shown in Fig.~\ref{fig:sdss_all}, the distributions look largely similar. According to the SDSS classification, none of these spectra belongs to O type whereas in our classification 185 spectra belong to that class. Similarly, there are 219 spectra which are labelled as B type in SDSS but according to our classification, 580 belong to that type. For a quantitative measure of the agreement between the two classifications, we find the average agreement of 60\% at the main-class level and 87\% after weighting each class with its population. 

\begin{figure*}
\centering
\includegraphics[scale=0.4]{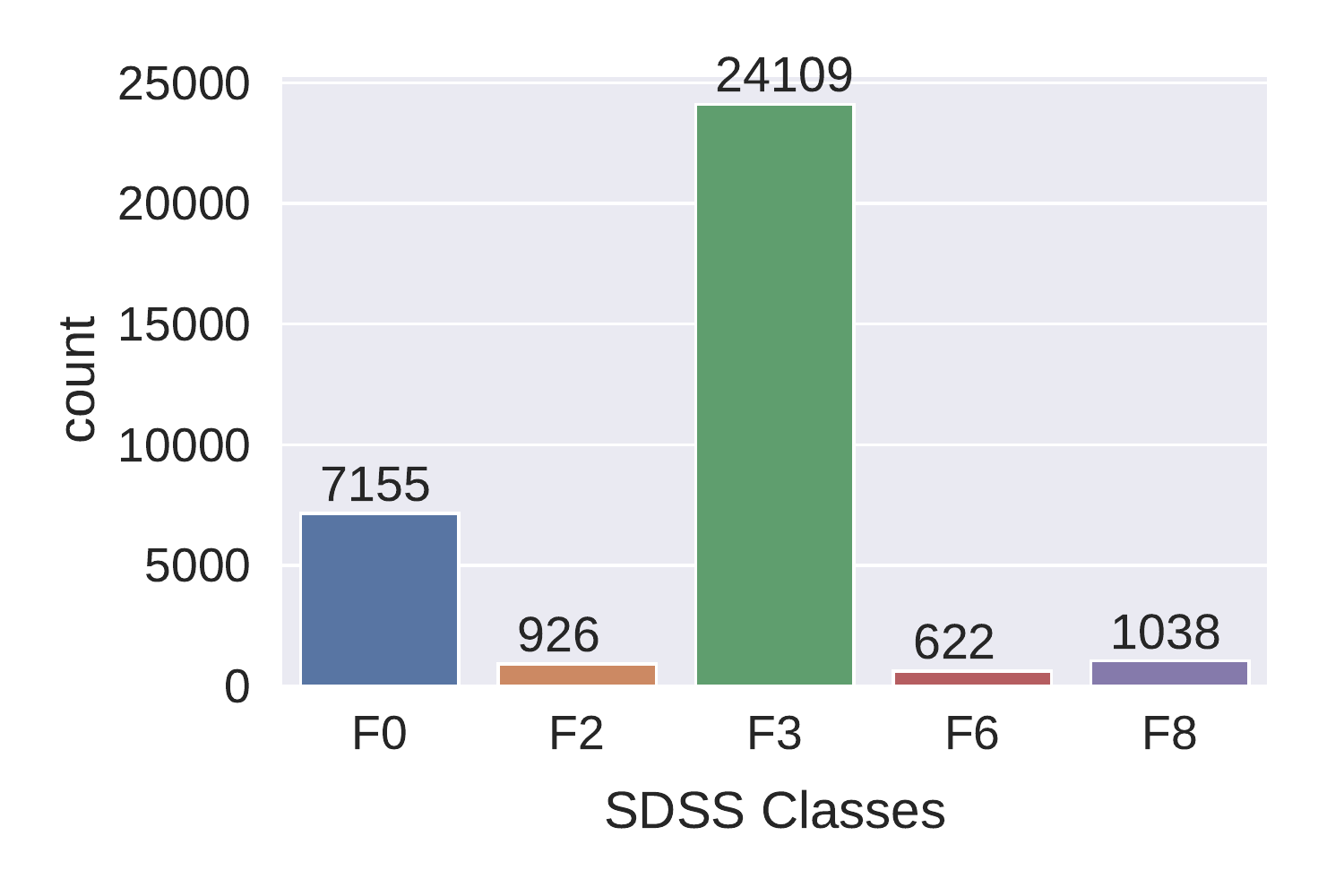}
\includegraphics[scale=0.4]{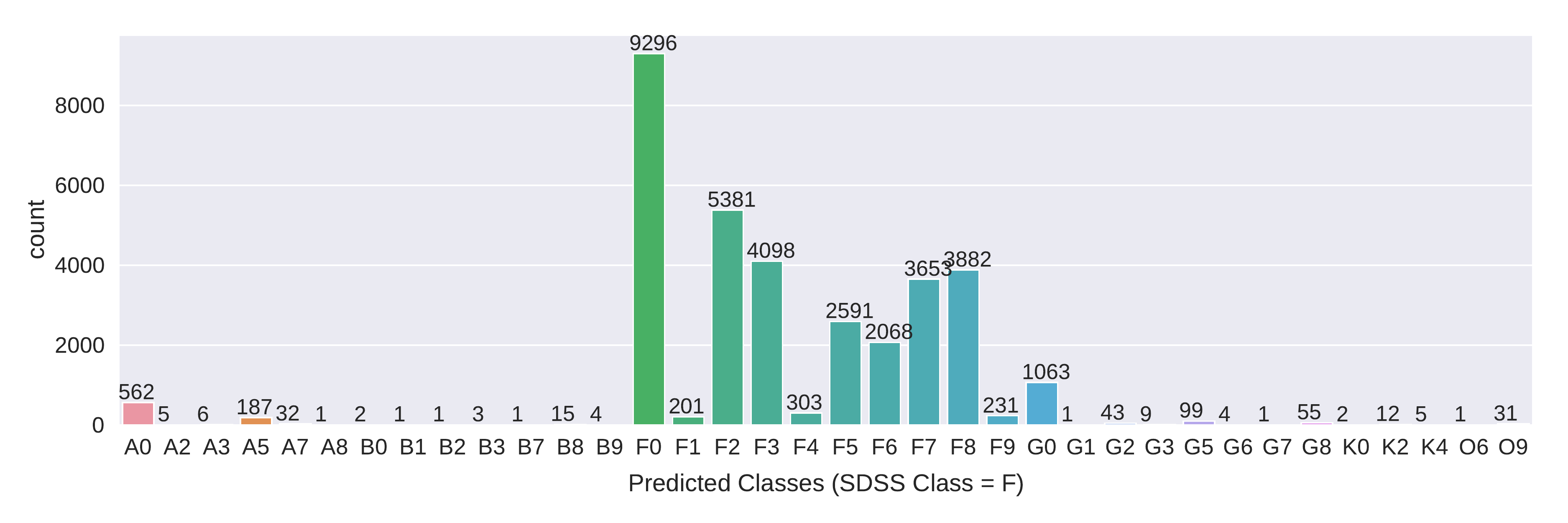}
\caption{Predicted class distribution of SDSS F type spectra. 
The upper panel shows the distribution of SDSS F type spectra into different subclasses as per the SDSS classifcation.
The lower panel presents the distribution of predicted classes for all F type spectra in SDSS.}
\label{fig:sdss_ftype}
\end{figure*}

For a deeper examination of the predicted classes, we check the classification at the spectral subclass level. At the subclass level, we get an average dispersion of $\sim$\,5 subclasses between our classification and the SDSS classification which comes down to $\sim$3 subclasses after removing 3$\sigma$ outliers iteratively. Since the SDSS sample is dominated by F type stars, we select all F type spectra in SDSS and compare the SDSS and predicted classifications. According to SDSS classification, 33945 F type spectra are distributed among the subclasses F0, F2, F3/5, F6 and F8 as 7155, 926, 24109, 622, and 1038 respectively. But our prediction suggests that SDSS F type sample contains about 800 spectra which are of A type and 1275 spectra belonging to the G class. The sample has very small contamination from the early type (B2, B3) as well as late type spectra (K4). The distribution of predicted classes for SDSS F type stars is presented in Fig.~\ref{fig:sdss_ftype}. 

\begin{figure*}
\centering
\includegraphics[scale=0.4]{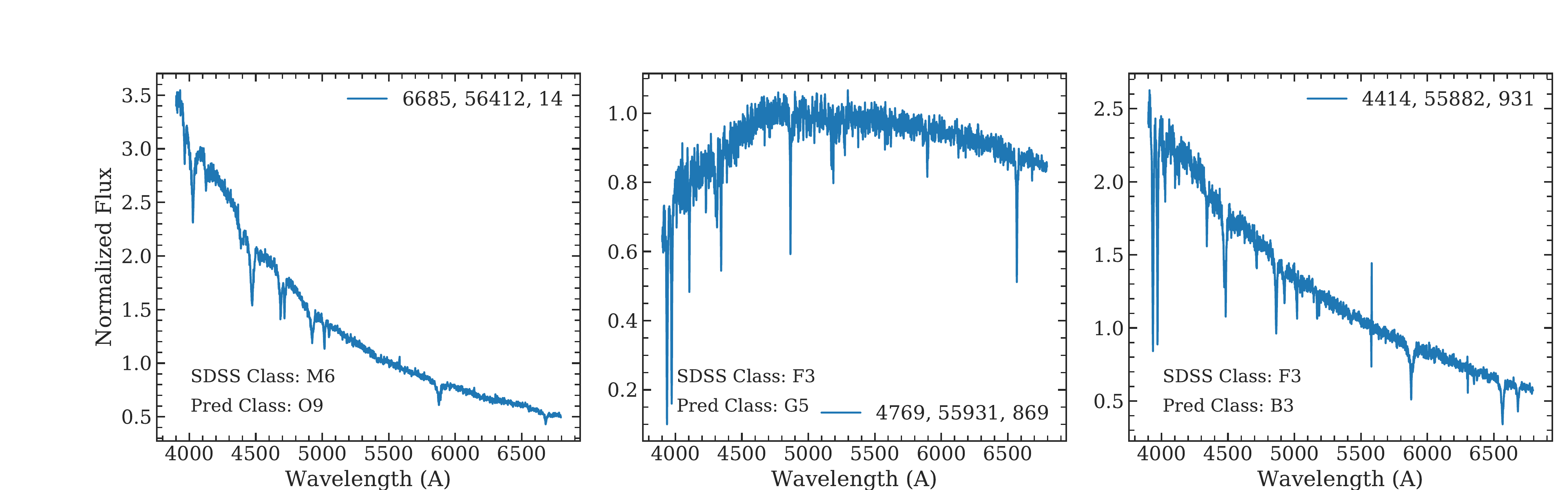}
\caption{A few sample SDSS Spectra for which the SDSS classification is in disagreement with our prediction. Legends in each panel indicate the SDSS identifiers in order of plate ID, MJD, and fiber ID respectively for the sample spectra.}
\label{fig:sdss_comparison}
\end{figure*}

Further investigating the differences between the SDSS and our classification by visual inspection of individual spectra, we find that in most cases, the difference between the two classification schemes is because of flux calibrations issues in SDSS spectra. However, there are a few cases where the assigned class by our classifier seems to be in better agreement with the spectral signatures and the SED. We show these examples in Fig.~\ref{fig:sdss_comparison}. The left panel in the figure shows an SDSS M6 type spectrum which is predicted as O9 by our classification model. This spectrum shows the typical spectral signatures of an O type star (He\,{\sevensize I} 3819, He\,{\sevensize I} 4026, He\,{\sevensize II} 4686 etc.), which agrees with our classification. Similarly, in the middle and right panel, we show the cases where SDSS pipeline assigns the spectra to the F3 class but our classification model predicts the spectra to be of G5 and B3 type respectively.

In Table~\ref{tab:sdss_classification}, we list the unique identifier of each individual SDSS spectrum based on three numbers: Plate, MJD, and Fiber ID. These three numbers combined serve as the spectrum identifier in SDSS. For each spectrum, we list the SDSS class and predicted class using our model.

\begin{table}
\caption{Expected and predicted classes for the SDSS spectra. The spectra are identified with a unique combination of plate, MJD and fiber ID. These three numbers together provide a unique identity to the SDSS spectra. Based on these three numbers, the SDSS spectral database can be queried. The table is published in its entirety in the electronic version. A portion is shown here for guidance regarding its form and content.}
\centering
\begin{tabular}{rrrll}
\hline\hline
 Plate &    MJD & Fiber ID &   SDSS Class & Predicted Class  \\
\hline
  6138 &  56598 &       934 &     F3/F5V &             F3V  \\
  6138 &  56598 &       716 &     F3/F5V &             F0V  \\
  6138 &  56598 &       547 &     F3/F5V &             F2V  \\
  7332 &  56683 &        94 &      K2III &           K2III  \\
  3692 &  55186 &       296 &        F8V &             G0V  \\
  7332 &  56683 &       186 &       K0IV &            K2IV  \\
  3615 &  55208 &        50 &     F3/F5V &             F2V  \\
  6425 &  56298 &       214 &     F3/F5V &             F0V  \\
  7332 &  56683 &       111 &       K0IV &             G9V  \\
  7332 &  56683 &       466 &       K0IV &             K2V  \\
\hline
\end{tabular}\label{tab:sdss_classification}
\end{table}

\section{Conclusions and discussion}\label{sec:results}

In this paper, we explore the use of advanced deep learning algorithms like CNN along with classical machine learning tools like ANN and Random Forest (RF) to address stellar spectral classification. ML methods like ANN and RF treat each flux value in a spectrum as an independent feature which is not the case with 1-D CNN, making them more suitable for such applications. In the 1-D CNN approach, we treat the flux values as a sequence of non-independent numbers and use convolutional kernels of various sizes to extract the relevant features along the wavelength axis.

We observe that the accuracy achieved using CNN is somewhat better than the traditional ML techniques. 
Deep CNN architectures require a large training sample for adjusting the weights of the network layers and to avoid over-fitting. Since our training sample is relatively small, we handle the situation using autoencoder architecture which functions as an unsupervised classifier and learns how to encode a spectrum to latent space representation and reconstruct it back from the embedding. Using the pre-trained encoder part with few more convolutional layers, we train a CNN-based stellar spectral classification model. Using this model, we achieve an average accuracy of 89\% for predicting main spectral classes. The predicted classes are accurate up to 1.23 subclasses. For the luminosity classification, we obtain an accuracy of 76\% using CNN with a classification error of 0.72 luminosity classes. In luminosity classification, we face the problem of an imbalanced training set because of the non availability of spectra of required luminosity class in the spectral libraries we use. For instance, compared to spectra belonging to luminosity class III and V, there were far fewer spectra in class I, II and IV. A better training sample would contain a comparable number of spectra in all the classes. While comparing our prediction of spectral classes with known classes for spectra in the CFLIB, we find that 15 spectra are misclassified to greater than five spectral subclasses. Studying these spectra in detail, we find that in most of the cases, our classification better conforms to the estimated parameters. Some of the misclassifications are a result of the absence of metal-poor stars (\feh{}\,<\,$-2.0$) in the training sample as a result of which all metal-poor stars are being classified to hotter classes. In some cases, the flux calibration was not accurate, leading to misclassification. We also demonstrate the capability of the algorithm to classify large stellar spectral databases by applying it to $\sim$ 48,000 SDSS spectra. 

While applying our classification model to CFLIB and SDSS spectra, we notice that our model is sensitive to the overall spectral energy distribution and provides an incorrect class if the flux calibration is not properly performed on the input spectrum. Sensitivity to the SED exists because the input spectra are primarily dominated by the SED and spectral features are relatively lesser. Therefore, the overall flux distribution influences the prediction of our classification model. Since large databases mostly rely on automatic data reduction pipelines which may sometimes fail in executing accurate calibrations, the sensitivity of the model to these calibrations might hinder its performance. We execute a few tests to evaluate the decline in the performance of the model in such cases and notice that the model can sustain against issues related to overall SED but with degraded performance. This shortcoming of the current model can be eliminated by training the network on normalized spectra where the continuum is normalized to unity everywhere. This training will be robust to the SED and the classification will be primarily based on the spectral lines. The major challenge with this approach is to prepare a large enough training sample of normalized spectra. The normalization process is more difficult for the later type spectra. Moreover, a model trained on normalized spectra will be applicable only to spectra which have already been normalized. Therefore, this will not be a suitable model for classifying spectra in flux-calibrated format from databases like SDSS and LAMOST. As a future scope of this work, we propose to integrate the two classification approaches, one based on flux calibrated pre-processed spectra and the other based on normalized spectra, into a single classification model which will be more robust to the faulty calibration issues. In the future, we also plan to use CNN with autoencoder architecture for finding rare and interesting astrophysical objects from large databases like SDSS, LAMOST, GALAH, etc using spectra to define extreme outliers. 

There are ongoing large efforts in assembling a list of benchmark stars, based on a holistic combination of analysis from high-resolution spectroscopy, photometry, astrometry and asteroseismology data \citep{Heiter2015,Strassmeier2018,Jofre2018}. Deep learning algorithms can take full advantage of these well defined stellar parameters of a large sample and can provide homogeneous and accurate stellar classification/parameters. This can be helpful in various astrophysical problems. For instance, identifying identical twin spectra among the benchmark samples with known distances, which can help us in climbing the distance ladder and provide distances to a large sample. Another application of such classification models would be to provide a simple number count of stars of mono-spectral type and luminosity class, which can help in probing the Galactic structure (e.g. possible flaring Galactic disk, stellar streams, thin/thick disk transition). Precision HST photometry has been instrumental in identifying the existence of multiple stellar populations in resolved star clusters \citep{Piotto2007,Anderson2009,Milone2010}. Well assembled spectral templates along with deep learning techniques can give accurate stellar parameters which can further help in uncovering multiple populations in globular clusters. These algorithms can also highlight different groups based on a statistically meaningful spread in the parameter space, which could be an indicator of other physical parameters not considered for the training e.g. stellar rotation, diffusion, spots, activity, age, etc. This will help in establishing correlations among different stellar parameters. 

\section*{Acknowledgements}

Ajit Kembhavi and Kaushal Sharma acknowledge financial support from a Raja Ramanna Fellowship. 

In this work, we have extensively used SDSS database. Funding for the Sloan Digital Sky Survey IV has been provided by the Alfred P. Sloan Foundation, the U.S. Department of Energy Office of Science, and the Participating Institutions. SDSS-IV acknowledges support and resources from the Center for High-Performance Computing at the University of Utah. The SDSS web site is \url{www.sdss.org}.

SDSS-IV is managed by the Astrophysical Research Consortium for the Participating Institutions of the SDSS Collaboration including the 
Brazilian Participation Group, the Carnegie Institution for Science, 
Carnegie Mellon University, the Chilean Participation Group, the French Participation Group, Harvard-Smithsonian Center for Astrophysics, 
Instituto de Astrof\'isica de Canarias, The Johns Hopkins University, Kavli Institute for the Physics and Mathematics of the Universe (IPMU) / 
University of Tokyo, the Korean Participation Group, Lawrence Berkeley National Laboratory, 
Leibniz Institut f\"ur Astrophysik Potsdam (AIP),  
Max-Planck-Institut f\"ur Astronomie (MPIA Heidelberg), 
Max-Planck-Institut f\"ur Astrophysik (MPA Garching), 
Max-Planck-Institut f\"ur Extraterrestrische Physik (MPE), 
National Astronomical Observatories of China, New Mexico State University, 
New York University, University of Notre Dame, 
Observat\'ario Nacional / MCTI, The Ohio State University, 
Pennsylvania State University, Shanghai Astronomical Observatory, 
United Kingdom Participation Group,
Universidad Nacional Aut\'onoma de M\'exico, University of Arizona, 
University of Colorado Boulder, University of Oxford, University of Portsmouth, 
University of Utah, University of Virginia, University of Washington, University of Wisconsin, 
Vanderbilt University, and Yale University.

\bibliographystyle{mnras}
\bibliography{bibliography}



\label{lastpage}

\end{document}